\documentclass[11pt,english,twoside]{article}

\usepackage[T1]{fontenc}
\usepackage[latin1]{inputenc}
\usepackage[english]{babel}
\usepackage{lmodern}

\usepackage{amsmath,amssymb}

\usepackage{float}
\usepackage{graphicx}
\usepackage[dvips]{epsfig}
\usepackage{psfrag}

\newcommand{\beq}{\begin{equation}}
\newcommand{\eeq}{\end{equation}}
\newcommand{\bea}{\begin{eqnarray}}
\newcommand{\eea}{\end{eqnarray}}
\newcommand{\nn}{\nonumber}
\newcommand{\w}{\wedge}
\newcommand{\sla}{\slash\!\!\!}

\begin{document}

\begin{titlepage}

\begin{flushright}

\end{flushright}

\vspace{.8in}

\begin{center}

{\Large \bf New supersymmetric flux vacua with intermediate $SU(2)$ structure}

\vspace{.8in}

{\Large David Andriot}

\vspace{.4in}

{LPTHE, CNRS, UPMC Univ Paris 06\\
Bo\^ite 126, 4 Place Jussieu\\
F-75252 Paris cedex 05, France}

\vspace{.2in}

{E-mail: \textit{andriot@lpthe.jussieu.fr}}

\end{center}

\vspace{.8in}

\begin{abstract}

\vspace{.2in}

We find new supersymmetric four-dimensional Minkowski flux vacua of type II string theory on nilmanifolds and solvmanifolds. We extend the results of M. Gra\~na, R. Minasian, M. Petrini, and A. Tomasiello to the case of intermediate $SU(2)$ structures (the two internal supersymmetry parameters are neither parallel nor orthogonal). As pointed out recently by P. Koerber and D. Tsimpis, intermediate $SU(2)$ structures are possible when one considers ``mixed'' orientifold projection conditions. To find our vacua, we rewrite these projection conditions in a more tractable way by introducing new variables: the projection basis. In these variables, the SUSY conditions become also much simpler to solve, and we find three new vacua. In addition, we find that these variables correspond to the $SU(2)$ structure appearing with the dielectric pure spinors, objects introduced and discussed by R. Minasian, M. Petrini, A. Zaffaroni, and N. Halmagyi, A. Tomasiello, in the AdS/CFT context. Besides, our solutions provide some intuition on what a dynamical $SU(3) \times SU(3)$ structure solution could look like.

\end{abstract}

\end{titlepage}

\tableofcontents

\newpage

\section{Introduction}

Flux compactifications \cite{FluxGrana} have appeared in the last few years as a promising approach to make contact between string theory and real world low energy physics. Indeed, considering non-trivial vacuum values for some supergravity fluxes on the internal manifold (on which one compactifies) has several interesting phenomenological consequences. For instance, one generates this way a potential in the effective quantum field theory, which lifts some of the moduli \cite{PS, TV}. One can also get a natural way to create hierarchies \cite{GKP}, and new possibilities for supersymmetry breaking \cite{PS, TV}.\\

Previously, in order to preserve the minimal amount of supersymmetry in the low energy effective theory, one was led to consider a Calabi-Yau (CY) as the compactifications manifold \cite{4Witt}. The introduction of background fluxes modifies the supersymmetry conditions, leading generically to new manifolds. The general mathematical characterization of these new manifolds was given in \cite{Gene04, Gene05}, where the authors rewrote the supersymmetry conditions in terms of Generalized Complex Geometry (GCG) \cite{Hitch, Gual}, and showed that the internal manifold has to be a (twisted) Generalized Calabi-Yau (GCY). An $\mathcal{N}=1$ supergravity vacuum generically needs the existence of a pair of two globally defined non-vanishing spinors on the internal manifold. A good object to characterize this pair is then the structure group on the tangent bundle $T$. Indeed, in six dimensions, this pair of spinors defines either an $SU(3)$ structure, a static $SU(2)$ structure, or what we will call here an intermediate $SU(2)$ structure, when respectively the spinors are parallel, orthogonal, or between the two. These different
possibilities are encoded, in the GCG context, into an $SU(3) \times SU(3)$ structure on the bundle $T \oplus T^*$. This structure is related in GCG to the existence of a pair of compatible pure spinors. When one of the two pure spinors is closed, the manifold is said to be a Generalized Calabi-Yau.\\

An interesting question is to find explicit examples of these new backreacted backgrounds. A successful approach \cite{KSTT} has been to start from a warped CY (in the simplest case a warped $T^6$) with an $O3$-plane\footnote{The compactification to four-dimensional Minkowski space-time needs the presence of space-filling orientifolds (O-planes) as sources in order to compensate the contribution of the fluxes to the energy-momentum tensor (the no-go theorem, or tadpole cancelation) \cite{GKP}.} and some background fluxes, and perform T-dualities to obtain new vacua on non-CY manifolds. In \cite{Scan}, the authors explored the possibility of using GCG to find ``new'' flux vacua, ``new'' in the sense they are neither conformal Calabi-Yau manifolds, nor T-dual to a warped $T^6$ with an $O3$: there are indeed some ``new'' vacua corresponding to nilmanifolds and solvmanifolds (twisted tori) with non trivial fluxes.\\

In their search for ``new'' four-dimensional Minkowski vacua, the authors of \cite{Scan} only looked for $SU(3)$ or static $SU(2)$ structures, since only those seemed to be compatible with the orientifold projection. Recently in \cite{KT}, it was shown that intermediate $SU(2)$ structures are also possible when one allows a mixing of the usual $SU(2)$ structure forms under the projection conditions. Then the last authors constructed such vacua on some GCY, starting from a warped $T^6$ with an $O3$ and performing some specific T-dualities.\\

In this paper, by first rewriting in a more tractable way the projection conditions imposed by the orientifold for intermediate $SU(2)$ structures, we manage to find for these structures genuinely ``new'' four-dimensional (Minkowski) flux vacua of type II string theory with (at least) $\mathcal{N}=1$. Note that we find them in the large volume limit with smeared sources, and for constant intermediate $SU(2)$ structures. These vacua are not T-dual to a warped $T^6$ with an $O3$ because the manifolds on which we find them, the same as in \cite{Scan}, do not have the right isometries to perform the needed T-dualities. Furthermore, by going to the limit in which the two internal spinors are parallel or orthogonal, we find back the solutions of \cite{Scan}, hence providing some idea of what a generic dynamical $SU(3) \times SU(3)$ structure should look like (a dynamical structure occurs when the internal spinors, hence the structure, are varying along the manifold).\\

The rewriting of the orientifold projection conditions is done by introducing what we call the projection (eigen)basis, i.e. the set of structure forms which are ``eigenvectors'' for the projection. These forms actually define a new $SU(2)$ structure, obtained by a rotation from the usual one. Moreover, we show that this $SU(2)$ structure is nothing (modulo a rescaling) but the one appearing with the dielectric pure spinors. The latter are a rewriting of the GCG pure spinors, used to study the deformations of four-dimensional $\mathcal{N}=4$ Super Yang-Mills in the context of AdS/CFT \cite{MPZ, Al}. As the pure spinors are much simpler when expressed with the projection basis variables, the supersymmetry conditions get much simpler. It is then easier to find solutions, which are nothing but the ``new'' vacua.\\

Here is how the paper is organized. In section \ref{secbckgd}, we give our supergravity conventions, the definitions of G-structures and our ansatz for the internal spinors, the GCG pure spinors and their properties, and finally a sum-up of the conditions a vacuum has to verify. In section \ref{secproj}, we derive as in \cite{KT} the projection conditions and rewrite them in a more tractable way by introducing the projection basis. Then we express the pure spinors in these variables, and explain the link with the dielectric pure spinors. Finally we give the SUSY conditions in these variables too. In section \ref{secsol}, after giving details on the set-up in which we are going to look for vacua, and the method used to find them, we give three solutions, among which two are T-duals. Then we study their limits to recover the solutions found in \cite{Scan}, and more. Finally, we look for other solutions in the specific case where there are several non completely overlapping orientifolds. In the appendix \ref{conv}, we give several conventions, a derivation of the $SU(2)$ structure conditions, and the proof that some of the structure conditions imply the compatibility conditions that should be verified by the pair of GCG pure spinors to define an $SU(3) \times SU(3)$ structure. In appendix \ref{Projbasap}, we give the structure conditions written in the projection basis variables, and details on the derivation of the SUSY conditions written in these variables too. In appendix \ref{cali}, we discuss some normalization condition related to the calibration of smeared sources. Details on the search for solutions with several orientifolds are given in appendix \ref{reducset}.

\section{Background}\label{secbckgd}

In this section, we give our supergravity conventions, discuss the parametrization of the internal spinors and their relation to the structure group. We also introduce pure spinors in GCG and give some of their properties. Finally, we formulate the conditions that a SUSY vacuum of type II string theory with fluxes has to satisfy, in terms of the pure spinors. Along this section, we mainly follow the conventions of \cite{Scan} and \cite{KT}. Some related details are given in appendix \ref{conv}.

\subsection{Supergravity conventions}\label{sugraconv}

In this paper we are interested in four-dimensional Minkowski flux vacua of type II string theory with (at least) $\mathcal {N}=1$ supersymmetry (SUSY).
Therefore we will consider type II supergravity (SUGRA) backgrounds, that are warped products of Minkowski $\mathbb{R}^{3,1}$ and of a six-dimensional compact space $M_6$ (assumed to be a smooth manifold). So we choose for these backgrounds the following metric:
\beq
\label{metricansatz}
ds_{(10)}^2=e^{2A(y)}\ \eta_{\mu\nu}dx^\mu dx^\nu
+g_{\mu\nu}(y) dy^\mu dy^\nu \ ,
\eeq
$\eta$ meaning here the diagonal Minkowski metric with signature $(-,+,+,+)$.
The solutions will also have non zero background values for some
of the RR and NS fluxes. Poincar\'e invariance in four dimensions requires the fluxes living on Minkowski to be proportional to $\textrm{vol}_{(4)}$, the warped four-dimensional volume form. So more interestingly, we will focus on non trivial fluxes living on the internal manifold.
As in \cite{Scan}, we define the total internal RR field $F$ as
\bea
\textrm{IIA}&:&\ F=F_0+F_2+F_4+F_6 \ , \label{RRIIA} \\
\textrm{IIB}&:&\ F=F_1+F_3+F_5 \ , \label{RRIIB}
\eea
with $F_k$ the internal $k$-form RR field. $F$ is related to the total ten-dimensional
RR field-strength $F^{(10)}$ by
\beq
F^{(10)}=F+\textrm{vol}_{(4)}\w \lambda(*F) \ ,
\eeq
where $*$ is the six-dimensional Hodge star, and $\lambda$ is an action defined on any $p$-form $A_p$ by a complete reversal of its indices
\beq
\lambda(A_p)=(-1)^{\frac{p(p-1)}{2}}A_p \label{lambda} \ .
\eeq

In order to find such solutions, one should solve the equations of motion and the Bianchi identities for
the fluxes. Actually, it has been proven in \cite{LT, GMSW, KT} that, for the class of supergravity backgrounds we are interested in, the equations of motion for the metric and the dilaton $\phi$ are implied by the Bianchi identities and the ten-dimensional supersymmetry conditions, so we will solve the latter. The ten-dimensional supersymmetry conditions are the annihilation of the supersymmetry variations of the gravitino $\psi_\mu$ and the dilatino $\lambda$, given by \cite{Gene04}
\bea
\delta \psi_\mu &=& D_\mu \epsilon + \frac{1}{4} H_\mu \mathcal{P} \epsilon + \frac{1}{16} e^{\phi} \sum_n \sla \!  {F_{2n}} \gamma_{\mu} \mathcal{P}_n \epsilon \ , \label{eq:susyg} \\
\delta \lambda &=& \left(\sla{\partial} \phi + \frac{1}{2} \sla \! H \mathcal{P}\right) \epsilon
+ \frac{1}{8} e^{\phi} \sum_n (-1)^{2n} (5-2n)\ \sla \! {F_{2n}} \mathcal{P}_n  \epsilon \ , \label{eq:susyd}
\eea
with $n=0,\ \ldots,\ 5$ for IIA and $n=\frac{1}{2},\ \ldots,\ \frac{9}{2}$ for IIB, and $H_\mu = \frac{1}{2}H_{\mu\nu\rho} \gamma^{\nu\rho}$, $H$ being the NSNS flux. The definitions of $\mathcal{P}$ and $\mathcal{P}_n$ are different in IIA and IIB: for IIA, $\mathcal{P} = \gamma_{11}$ and $\mathcal{P}_n = \gamma_{11}^n \sigma^1$, while for
IIB,  $\mathcal{P} = -\sigma^3$, $\mathcal{P}_n = \sigma^1$ for $n+\frac{1}{2}$ even and $\mathcal{P}_n = i \sigma^2$ for $n+\frac{1}{2}$ odd. The two Majorana-Weyl supersymmetry parameters of type II supergravity are
arranged in the doublet $\epsilon= (\epsilon^1,\epsilon^2)$.\\

Because of the product structure of the solution (\ref{metricansatz}), the Lorentz group is broken to $SO(1,3) \times SO(6)$ and the supersymmetry
parameters $\epsilon^i$ should be decomposed accordingly. This means there should be on the compact manifold $M_6$ a set of independent globally defined and non-vanishing spinors noted $\eta^{i}_{a}$ on which one can expand the $\epsilon^i$ as
\bea
\label{10dspinoransatz}
\epsilon^1 &=& \zeta^1 \otimes \sum_{a} \alpha^1_a \eta^1_a + c.c. \ , \nn\\
\epsilon^2 &=& \zeta^2 \otimes \sum_{a} \alpha^2_a \eta^2_a + c.c. \ .
\eea

In this formulation, the $\zeta^i$ are the four-dimensional SUSY parameters, and the decomposition on the six-dimensional (internal) spinors can be seen from the four-dimensional point of view as internal degrees of freedom of the $\zeta^i$. Hence, the number $\mathcal{N}$ of four-dimensional SUSYs is increased by one for each non-zero $\alpha^i_a$ with the corresponding internal spinor $\eta^i_a$ being a Killing spinor for the SUSY conditions. So, to get at least a $\mathcal{N}=1$ vacuum as we want, one needs at least a pair $(\eta^1,\eta^2)$ of globally defined non-vanishing internal spinors that satisfy the SUSY conditions (and for $\mathcal{N}=1$ one also needs $\zeta^1=\zeta^2$). Let us now see how to parametrize this pair of internal spinors, and their relations with the G-structures one can define on the manifold.

\subsection{Internal spinors and G-structures variables}\label{strucgpprop}

A manifold $M$ is said to admit a G-structure when its structure group is reduced to
the subgroup G. The reduction is associated to the existence on the manifold of
globally defined spinors. Here we are interested in $SU(3)$ and $SU(2)$ structures in six-dimensions.\\

An $SU(3)$ structure is defined by a globally defined non-vanishing spinor $\eta_+$.
In six dimensions, this spinor is a Weyl spinor so it has definite chirality. Here we take $\eta_+$ of positive chirality and of unitary norm. Complex conjugation acts as $(\eta_+)^*=\eta_-$.
A G-structure is equivalently defined in terms of G-invariant no-where
vanishing globally defined forms. These can be obtained as bilinears of the
globally  defined spinors. For an $SU(3)$ structure, one can define a holomorphic three-form $\Omega_3$
and a K\"ahler form $J$ given by\footnote{\label{holo}For both $SU(3)$ and $SU(2)$ structures, the holomorphicity of forms is defined with respect to the almost complex structure given in subsection \ref{Solconv}. The indices $\mu,\ \nu,\ \rho$ are real.}
\bea
\label{SU3forms}
\Omega_{\mu\nu\rho} &=& - i \eta_-^{\dag}\gamma_{\mu\nu\rho}\eta_+ \ , \nn \\
J_{\mu\nu} &=& - i \eta_+^{\dag}\gamma_{\mu\nu}\eta_+  \ ,
\eea
satisfying the structure conditions
\beq
\label{SU3comp}
J\w \Omega_3=0 \qquad \frac{4}{3}J^3= i \Omega_3\w \overline{\Omega}_3 \neq 0  \ .
\eeq

Similarly, an $SU(2)$ structure is defined by two orthogonal globally defined spinors $\eta_+$ and $\chi_+$ (we take them of unitary norm). In terms of invariant forms, an $SU(2)$
structure is given by a holomorphic one-form $z$ (we take $||z||^2=2$), a real two-form $j$ and
a holomorphic two-form $\Omega_2$ given by
\bea
\label{SU2forms}
z_\mu &=& \eta_-^{\dagger} \gamma_\mu \chi_+ \ , \nn \\
j_{\mu\nu} &=& - i \eta_+^{\dag}\gamma_{\mu\nu}\eta_+ +i \chi_+^{\dag}\gamma_{\mu\nu}\chi_+ \ , \nn \\
\Omega_{\mu\nu} &=& \eta_-^{\dagger} \gamma_{\mu\nu} \chi_- \ ,
\eea
satisfying the following structure conditions
\bea
&&\!\!\!\!\!\!\!\!\!\!\!\!\!\!\!\!\!\!\!\!\!\!\!\!\!\! j^2=\frac{1}{2}\ \Omega_2\w \overline{\Omega}_2 \neq 0 \ , \label{SU2comp1} \\
j\w \Omega_2=0\ ,&& \quad \Omega_2\w \Omega_2=0 \ , \label{SU2comp2} \\
z \llcorner \Omega_2 = 0\ ,&& \quad z \llcorner j = 0 \ .\label{SU2comp3}
\eea
where the definition of the contraction $\llcorner$ is given in appendix \ref{Convdiff}. We give one possible derivation of these structure conditions in appendix \ref{SU(2)}.\\

Note that it is possible to rewrite the spinor $\chi_+$ as
\beq
\chi_+ = \frac{1}{2} z \eta_- \ .
\eeq

The $SU(2)$ structure is naturally embedded in the $SU(3)$ structure defined by $\eta_+$:
\beq
J=j+\frac{i}{2}z\w \overline{z} \ , \qquad \Omega_3= z\w \Omega_2 \ , \label{SU2}
\eeq
and one then has the reverse relations
\beq
j=J-\frac{i}{2}z\w \overline{z} \ , \qquad \Omega_2=\frac{1}{2}\overline{z}\llcorner \Omega_3 \ .
\eeq

Let us consider now a pair of globally defined non-vanishing internal spinors,
$\eta^{1}_+$ and $\eta^{2}_+$, corresponding to the internal components of the
supersymmetry parameters. We choose to parametrize them this way (always possible):
\bea
\label{defk}
\eta^1_+ &=& a \eta_+ \ , \nn \\
\eta^2_+ &=& b (k_{||}\eta_+ +k_{\bot} \frac{z\eta_-}{2} ) \ .
\eea
$\eta_+$ and $\chi_+= \frac{1}{2} z \eta_-$ in (\ref{defk}) define an $SU(2)$ structure in the way explained before.
$k_{||}$ is real and $0\leq k_{||} \leq 1$, $k_{\bot}=\sqrt{1-k_{||}^2}$. $a$ and $b$ are never-vanishing complex numbers related to the norms of the
spinors $\eta^{i}_+$:
\beq
||\eta_+^1||= |a| \ , \qquad ||\eta_+^2||= |b| \ . \label{normseta}
\eeq
In the rest of the paper we will always take $|a|=|b|$, so that $||\eta_+^1||=||\eta_+^2||$.
As we will see later, this condition is implied by the orientifold projection. The relative phases of the spinors
can be fixed by introducing $e^{i\theta}=\frac{b}{a}$ and imposing $b=\overline{a}$. The remaining freedom is then only in $\theta$ and $|a|$.\\

Depending on the values of the parameters $k_{||}$ and $k_{\bot}$, one can define from these spinors different
G-structures on the internal manifold. If one takes $k_{\bot}=0$, the $\eta_+^i$ become parallel, hence there is only one globally defined non-vanishing spinor, and this corresponds to an $SU(3)$ structure.
When $k_{\bot}\neq0$, the two spinors are genuinely independent,
and so we get an $SU(2)$ structure \cite{DallAg}. In the particular case $k_{||}=0$, i.e. $k_{\bot}=1$,
the spinors are  orthogonal, and this corresponds to what is called in the literature
a static $SU(2)$ structure. In the intermediate case ($k_{||}\neq0$, $k_{\bot}\neq0$),
we have what is sometimes called a dynamical $SU(2)$ structure, in reference to the fact these coefficients
could change when we move on the manifold. We prefer to call it an intermediate $SU(2)$
structure, because these coefficients can also be constant but still non-zero
(and then the structure is not properly speaking dynamical).\\

It is clear that $k_{||}$ and $k_{\bot}$ can be related to the ``angle'' between the spinors. We can introduce the angle $\phi$
\beq
k_{||}=\cos(\phi),\ k_{\bot}=\sin(\phi),\ 0\leq\phi\leq\frac{\pi}{2} \ , \label{kparam}
\eeq
and we get the following pictures of the different structures:
\begin{figure}[H]
\begin{center}
\begin{tabular}{ccccc}

\psfrag{eta1}[][b]{$\eta_+^1$}
\psfrag{eta2}[l][]{$\eta_+^2$}
\includegraphics{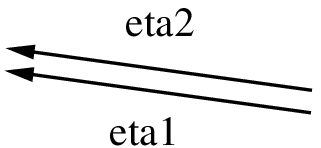}&

 &

\psfrag{eta1}[][]{$\eta_+^1$}
\psfrag{eta2}[][]{$\eta_+^2$}
\psfrag{Phi}[][]{$\phi$}
\includegraphics{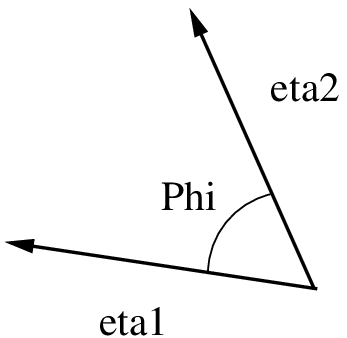}&

 &

\psfrag{eta1}[l][]{$\eta_+^1$}
\psfrag{eta2}[][]{$\eta_+^2$}
\includegraphics{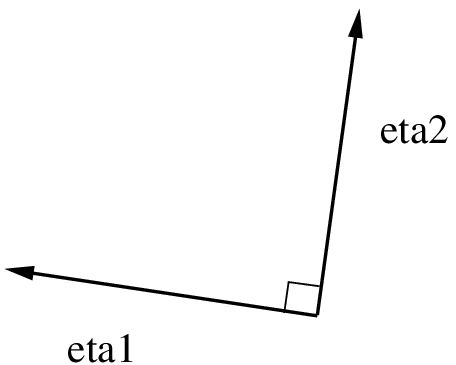}\\

 & & & & \\

$SU(3)$ structure:&

 &

Intermediate $SU(2)$ structure:&

 &

Static $SU(2)$ structure:\\

$k_{||}=1,\ k_{\bot}=0$&

 &

$k_{||}\neq0,\ k_{\bot}\neq0$&

 &

$k_{||}=0,\ k_{\bot}=1$

\end{tabular}\caption{The different structures}\label{figstruc}
\end{center}
\end{figure}

As a comparison to (\ref{SU2}), one can work out the embedding of the defined $SU(2)$ structure in the $SU(3)$ structure defined by $\frac{\eta^2_+}{||\eta^2_+||}$ ($\tilde J$ and $\tilde \Omega_3$). It is given by the previous $U(1)$ parameter $\phi$ \cite{DallAg}:
\bea
\tilde J&=&\cos(2\phi) j +\frac{i}{2} z\w \overline{z} + \sin(2\phi)\textrm{Re}(\Omega_2) \ ,\\
\tilde \Omega_3&=&-\sin(2\phi) z\w j + z\w (\cos(2\phi)\textrm{Re}(\Omega_2)+i\textrm{Im}(\Omega_2)) \ .
\eea

\subsection{Pure spinors of GCG and properties}\label{Puretype}

To solve the SUSY conditions, rather than using Killing spinors methods or G-structures tools, we will use the formalism of Generalized Complex Geometry (GCG). In Generalized Complex Geometry, given a manifold $M_d$ of real dimension $d$, one considers the bundle $T \oplus T^*$, whose sections are generalized vectors (sums of a vector and a $1$-form). For a review on GCG, see for instance \cite{Scan} or the original works \cite{Hitch}
and \cite{Gual}. In this paper we will be interested in the spinors on $T \oplus T^*$. These are Majorana-Weyl $\textrm{Cliff}(d,d)$ spinors, and locally they can be seen
as polyforms: sums of even/odd differential forms, which correspond to positive/negative chirality spinors.
A $\textrm{Cliff}(d,d)$ spinor is pure if it is annihilated by half of the $\textrm{Cliff}(d,d)$ gamma matrices. Such pure spinors can be obtained as tensor products of $\textrm{Cliff}(d)$ spinors, since
bispinors are isomorphic to forms via the Clifford map
\beq
C = \sum_k \frac{1}{k!} C^{(k)}_{i_1 \ldots i_k} dx^{i_1} \w \dots \w dx^{i_k}
\quad \leftrightarrow \quad C = \sum_k \frac{1}{k!} C^{(k)}_{i_1 \ldots i_k}
\gamma^{i_1 \dots i_k} \ .
\eeq

In the supergravity context, it is therefore natural to define the $\textrm{Cliff}(6,6)$ pure spinors as a bi-product of the
internal supersymmetry parameters
\bea
\label{purespinordef}
\Phi_+ &=& \eta^{1}_+ \otimes \eta^{2\dag}_+ \ , \nn \\
\Phi_- &=& \eta^{1}_+ \otimes \eta^{2\dag}_-  \ .
\eea
They can be seen as polyforms via the Fierz identity
\beq
\eta^{1}_+ \otimes \eta^{2\dag}_\pm=\frac{1}{8}\sum_{k=0}^6 \frac{1}{k!}\left( \eta^{2\dag}_\pm \gamma_{\mu_k...\mu_1} \eta^{1}_+ \right ) \gamma^{\mu_1...\mu_k} \ .
\eeq
The explicit expressions of the two pure spinors can then obtained \cite{JW} using the definitions of last subsection
\bea
\label{purespin}
\Phi_+ &=& \frac{|a|^2}{8} e^{- i \theta} e^{\frac{1}{2} z\w \overline{z}} (k_{||}e^{-ij}-ik_{\bot}\Omega_2) \ , \nn \\
\Phi_- &=& -\frac{|a|^2}{8} z\w (k_{\bot}e^{-ij}+i k_{||}\Omega_2) \ .
\eea

A pure spinor $\Psi$ can always be written as \cite{Gual}
\beq
\Psi=\Omega_k \w e^{B + i \omega}
\eeq
where $\Omega_k$ is a holomorphic $k$-form, and $B$ and $\omega$ are real
two-forms. The rank $k$ of $\Omega_k$ is called the type of the spinor. For the intermediate $SU(2)$ structure where both $k_{||}$ and $k_{\bot}$ are non zero,
it is possible to ``exponentiate'' $\Omega_2$ and get from (\ref{purespin})
\bea
\label{type}
\Phi_+ &=& \frac{|a|^2}{8}e^{-i\theta}k_{||}\ e^{\frac{1}{2} z\w \overline{z}-ij-i\frac{k_{\bot}}{k_{||}}\Omega_2} \ , \nn \\
\Phi_- &=& -\frac{|a|^2}{8}k_{\bot}\ z\w e^{-ij+i\frac{k_{||}}{k_{\bot}}\Omega_2} \ ,
\eea
so that the spinors have definite types: 0 and 1. In the case of the $SU(3)$ structure limit ($k_{\bot}=0$), we get that pure spinors are of type 0 and 3
\beq
\Phi_+ = \frac{|a|^2}{8} e^{-i \theta} e^{-iJ} \ , \qquad \Phi_- = - i \frac{|a|^2 }{8}\Omega_3 \ ,
\eeq
while in the case of the other limit, the static $SU(2)$ structure ($k_{||}=0$), the types are 1 and 2:
\beq
\Phi_+ = -i \frac{|a|^2}{8} e^{-i \theta} \Omega_2 \w e^{\frac{1}{2} z \w \bar{z}} \ , \qquad \Phi_- = - \frac{|a|^2 }{8} z \w e^{-i j} \ .
\eeq

Two pure spinors are said to be compatible if they have three common annihilators. This
can be rephrased in a set of compatibility conditions the spinors must satisfy.
We introduce the Mukai pairing for two polyforms $\Psi_i$:
\beq
\left\langle \Psi_1, \Psi_2 \right\rangle =\left( \Psi_1\w \lambda(\Psi_2) \right)_{\textrm{top}} \ , \label{Mukai}
\eeq
where $\textrm{top}$ means the top-form, and $\lambda$ has been defined in (\ref{lambda}).
It is also useful to recall the action of a generalized vector
$X=(x,y)\ \in\ T \oplus T^*$ on a polyform
\beq
X \cdot \Psi_i=x \llcorner \Psi_i + y \w \Psi_i \ .\label{Xaction}
\eeq
Then the compatibility conditions of two pure spinors $\Phi_1$ and $\Phi_2$ read
\bea
\label{compcond}
&& \left\langle \Phi_1, \overline{\Phi}_1  \right\rangle=\left\langle \Phi_2, \overline{\Phi}_2  \right\rangle \neq 0 \ , \\
&& \left\langle \Phi_1, X \cdot \Phi_2 \right\rangle=\left\langle \overline{\Phi}_1 , X \cdot \Phi_2 \right\rangle=0,\ \forall\ X\ \in\ T \oplus T^* \ .
\eea
A pair of compatible pure spinors defines an $SU(3) \times SU(3)$ structure on $T \oplus T^*$.
Depending on the relation between the spinors $\eta^{1,2}_+$, this translates on $T$ into the $SU(3)$,
static $SU(2)$ or intermediate $SU(2)$ structures discussed above. So the formalism of GCG allows to give a
unified characterization of the topological properties a $\mathcal{N}=1$ vacuum has to satisfy:
it must admit an $SU(3) \times SU(3)$ structure on $T \oplus T^*$. And so to satisfy this condition, we will verify that our vacua admit a pair of compatible pure spinors. One can actually show (see appendix \ref{proofcom}) that the ``wedge'' structure conditions (\ref{SU3comp}), or (\ref{SU2comp1}) and (\ref{SU2comp2}), imply the compatibility conditions in any of the three cases, so one can verify that these conditions are satisfied, instead of the compatibility ones.

\subsection{Conditions for a SUSY vacuum}\label{cond}

An $\mathcal{N}=1$ vacuum described in subsection \ref{sugraconv} should satisfy the SUSY conditions, the equations of motion (e.o.m.) and the Bianchi identities (BI) for the fluxes. In \cite{Gene04}, the SUSY conditions given in (\ref{eq:susyg}) and (\ref{eq:susyd}) were rewritten as differential conditions on the pure spinors:
\bea
&& (d-H\w)(e^{2A-\phi}\Phi_1)=0 \label{susy1} \ , \\
&& (d-H\w)(e^{A-\phi}\textrm{Re}(\Phi_2))=0 \label{susy2} \ , \\
&& (d-H\w)(e^{3A-\phi}\textrm{Im}(\Phi_2))=\frac{e^{4A}}{8}*\lambda(F) \ , \label{susy3}
\eea
with $\lambda$ defined in (\ref{lambda}), and with
\beq
\Phi_1=\Phi_\pm \qquad \Phi_2=\Phi_\mp \label{Phi12pm} \ ,
\eeq
for IIA/IIB (upper/lower) (conventions of \cite{Scan}). These conditions generalize the Calabi-Yau condition for fluxless compactifications. Indeed, the first of these equations implies that one of the two pure spinors (the one with the same parity as the RR fields) must be twisted (because of the $-H\w$) conformally closed. A manifold admitting a twisted closed pure spinor is a twisted Generalized Calabi-Yau (GCY, see the precise definition in \cite{Hitch, Gual} or \cite{Scan}). So we will look for vacua on such manifolds.\\

The e.o.m of the fluxes read
\beq
(d+H\w)(e^{4A}*F)=0 \ , \qquad d(e^{4A-2\phi}*H)=\mp e^{4A} \sum_n F_n \w * F_{n+2} \ , \label{eom}
\eeq
with the upper/lower sign for IIA/IIB. The BI (we assume no NS source) are
\beq
(d - H \w ) F=\delta(source) \ , \qquad dH=0 \ . \label{BI}
\eeq
Here $\delta(source)$ is the charge density of the allowed sources: these are space-filling
D-branes or orientifold planes (O-planes). In compactification to four-dimensional Minkowski, the trace of the energy-momentum tensor must be zero. This is the tadpole cancelation condition or no-go theorem \cite{GKP}. Then O-planes are needed since they are the only known sources with a negative charge, that can therefore cancel the flux contribution to this trace. As in \cite{Scan}, in this paper we will consider smeared sources, i.e. the sources are not localized anymore. The RR BI are then assumed to be
\beq
(d - H \w )F =\sum_{i} Q_i V^i \ ,\label{BI2}
\eeq
where $Q_i$ is the source charge and $V^i$ is (up to a sign) its internal co-volume (the co-volume of the cycle wrapped by the source). The sign of the $Q_i$ indicates whether the source is a D-brane ($Q_i > 0$) or an O-plane ($Q_i < 0$). For more details, see section \ref{secsol} and appendix \ref{cali}.\\

For intermediate $SU(2)$ structures (for which $\frac{k_{\bot}}{k_{||}}$ is constant) in the large volume limit (see subsection \ref{SUSYsec}), we will get from our SUSY conditions ((\ref{SUSYapA}) and (\ref{SUSYapB})) that the $H$ BI is automatically satisfied. Furthermore, for this class of compactifications, it was shown in \cite{Scan} that the e.o.m. for the RR fluxes are implied by the SUSY conditions. And it was shown in \cite{KT} that the e.o.m. of $H$ is implied by the SUSY conditions and the BI. So to sum-up, in order to find a solution, having a pair of compatible pure spinors on an GCY with at least one O-plane, we will have to verify that the SUSY conditions and the RR BI are satisfied.

\section{Projection conditions and consequences}\label{secproj}

As discussed in the previous subsection, tadpole cancelation requires the inclusion in the
solutions of O-plane sources. The presence of O-planes implies that the solution has to be invariant under the action of the orientifold. This imposes some projection conditions on the fields: one has to mod out by $ \Omega_{WS} (-1)^{F_L} \sigma$ for $O3/O7$ and $O6$, and by
$ \Omega_{WS} \sigma$ for $O5/O9$ and $O4/O8$. $\Omega_{WS}$ is a world-sheet reflection, $F_L$
is the left-movers fermion number, and $\sigma$ is an involution on the target space. The orientifold action on the pure spinors for $SU(3) \times SU(3)$ manifold were worked-out in \cite{Scan} (see also \cite{Grimm}). The authors of \cite{Scan} concluded that the orientifold projections are only compatible with
$SU(3)$ or static $SU(2)$ structures. Actually, as shown in \cite{KT}, intermediate $SU(2)$ structures are also compatible with O5-, O6- and O7-planes, if one allows a mixing between the two-forms specifying the structure. In this paper, we will only consider O5- and O6-planes.\\

In the first subsection we repeat the derivation of the orientifold projection conditions of \cite{KT} for O5- and O6-planes. The resulting conditions on the $SU(2)$ structure forms ($j$, $\Omega_2$ and $z$) appear to be not very tractable. We then show in the following subsection that it is possible to rewrite these conditions in a more tractable manner, which will allow us to find directly solutions. To do so, we introduce the projection (eigen)basis, and then write the pure spinors in these variables, and discuss their relation to the dielectric ones \cite{MPZ, Al}. Finally, we also give the supersymmetry conditions in the projection basis (details on the derivation are in appendix \ref{SUSYapprincipal}), and do the same for some structure conditions in appendix \ref{projbasis}.

\subsection{The orientifold projection}

As shown in \cite{KT}, the first step to derive the orientifold projection on the pure spinors
is to compute those for the internal SUSY parameters. This can be done starting from
the projection on the ten-dimensional SUSY spinorial parameters $\epsilon^{i}$,
and then reducing to the internal spinors $\eta^{i}_\pm$. In our conventions, we get
\bea
O5&:& \sigma(\eta_{\pm}^1)=\eta_{\pm}^2 \qquad \sigma(\eta_{\pm}^2)=\eta_{\pm}^1 \label{projeta1} \ , \\
O6&:& \sigma(\eta_{\pm}^1)=\eta_{\mp}^2 \qquad \sigma(\eta_{\pm}^2)=\eta_{\mp}^1 \label{projeta2} \ .
\eea
$\sigma$ is the target space reflection in the directions transverse to the O-plane. Using the expressions for the internal spinors given in (\ref{defk}), we obtain the following projection conditions at the orientifold plane:
\bea
O5&:& e^{i\theta}=\pm1,\ z\perp O5 \label{O5t} \ , \\
O6&:& e^{i\theta}\ \textrm{free},\ \textrm{Re}(z)\parallel O6,\ \textrm{Im}(z)\perp O6 \ . \label{O6t}
\eea
We can reexpress the previous conditions on $z$ in the following way:
\bea
\label{proj0}
O5&:& \sigma(z)=-z \ , \nn \\
O6&:& \sigma(z)=\overline{z} \ .
\eea
As explained in \cite{KT}, if the G-structures considered are constant (we will assume so), and if we work on nil/solvmanifolds (which will be our case), these conditions are valid everywhere (not only at the orientifold plane).\\

Following \cite{KT}, starting from the projections on the $\eta^{i}_\pm$, we derive
the projections of the pure spinors $\Phi_\pm$, and from them those for
the $SU(2)$ structure forms (using (\ref{O5t}) and (\ref{O6t})). To do this last step, one has to know that,
as $\sigma$ is only the reflection due to the orientifold, it can distributed on every term
of a wedge product. Furthermore, $\lambda(..)$ can also be distributed on wedge products of two forms, provided that one of the two forms is even (see (\ref{lambda2})). So we recover the same projection conditions on the forms as they have in \cite{KT}\footnote{We use slightly different conventions than in \cite{KT} but actually one can start with the following general expressions which cover both articles' conventions:
\bea
\Phi_+ &=& \frac{|a|^2}{8}e^{-i\theta}N^2 e^{\frac{1}{||z||^2} z\w \overline{z}} (k_{||}e^{-ij}-ik_{\bot}\Omega_2) \ , \nn \\
\Phi_- &=& -\frac{|a|^2}{8}\frac{\sqrt{2}}{||z||}N^2 z\w (k_{\bot}e^{-ij}+i k_{||}\Omega_2) \ , \nn
\eea
with $|a|,\ \theta,\ ||z||,\ ||\eta_+||=N$ constant and non-zero, and $k_{||},\ k_{\bot}$ constant, and then one gets the same projection conditions.}:
\bea
O5&:& \sigma(j)=(k_{||}^2-k_{\bot}^2)j+2k_{||}k_{\bot}\textrm{Re}(\Omega_2) \ , \nn \\
& & \sigma(\Omega_2)=-k_{||}^2\Omega_2+k_{\bot}^2\overline{\Omega}_2 +2k_{||}k_{\bot}j \ , \\
O6&:& \sigma(j)=-(k_{||}^2-k_{\bot}^2)j-2k_{||}k_{\bot}\textrm{Re}(\Omega_2) \ , \nn \\
& & \sigma(\Omega_2)=k_{||}^2\overline{\Omega}_2 -k_{\bot}^2\Omega_2-2k_{||}k_{\bot}j \ .
\eea
By introducing as in \cite{KT}:
\bea
O5&:&  k_{||}=\cos(\phi),\ k_{\bot}=\sin(\phi),\ 0\leq\phi\leq\frac{\pi}{2} \ , \\
O6&:&  k_{||}=\cos(\phi+\frac{\pi}{2})=-\sin(\phi), \nn \\
& & k_{\bot}=\sin(\phi+\frac{\pi}{2})=\cos(\phi),\ -\frac{\pi}{2}\leq\phi\leq0 \label{kphi} \ ,
\eea
we get in both cases the more convenient formulas:
\bea
\sigma(j) &=& \cos(2\phi) j+ \sin(2\phi) \textrm{Re}(\Omega_2) \ , \nn\\
\sigma(\textrm{Re}(\Omega_2)) &=& \sin(2\phi)j - \cos(2\phi) \textrm{Re}(\Omega_2) \ , \nn\\
\sigma(\textrm{Im}(\Omega_2)) &=& - \textrm{Im}(\Omega_2) \label{proj1} \ .
\eea

\subsection{The projection basis}

If one is looking for solutions to the projection conditions (\ref{proj1}), one will notice that they are
not very tractable. A good idea is to work in the projection (eigen)basis:
\bea
j_{||}=\frac{1}{2} (j+\sigma(j))\ , && j_{\bot}=\frac{1}{2} (j-\sigma(j)) \ , \nn\\
\textrm{Re}(\Omega_2)_{||}=\frac{1}{2}(\textrm{Re}(\Omega_2)+\sigma(\textrm{Re}(\Omega_2)))\ , && \textrm{Re}(\Omega_2)_{\bot}=\frac{1}{2} (\textrm{Re}(\Omega_2)-\sigma(\textrm{Re}(\Omega_2))) \ .
\eea
Using the property $\sigma^2=1$ and applying it to the previous equations, we get these more tractable equations:
\bea
&& j_{||}\ (1-\cos(2\phi))=\sin(2\phi)\ \textrm{Re}(\Omega_2)_{||} \ , \nn\\
&& j_{\bot}\ (1+\cos(2\phi))=-\sin(2\phi)\ \textrm{Re}(\Omega_2)_{\bot} \ .\label{proj2}
\eea
We also get the following equations:
\bea
&& j_{||}\ \sin(2\phi)=(1+\cos(2\phi))\ \textrm{Re}(\Omega_2)_{||} \ , \nn\\
&& j_{\bot}\ \sin(2\phi)= - (1-\cos(2\phi))\ \textrm{Re}(\Omega_2)_{\bot} \ ,
\eea
which are equivalent to the two equations (\ref{proj2}) if $k_{||}$ and $k_{\bot}$ are non-zero. It will be our case, so we will not use them. If we introduce (assuming that $k_{||}$ and $k_{\bot}$ are non-zero)
\bea
O5&:& \gamma=\frac{k_{||}}{k_{\bot}} \ , \nn \\
O6&:& \gamma=-\frac{k_{\bot}}{k_{||}} \ ,
\eea
and $r=\mp1$ for $O6/O5$ (upper/lower), the projection conditions become for both theories:
\bea
\label{projb1}
\sigma(\textrm{Re}(z))&=&-r\ \textrm{Re}(z) \ , \nn\\
\sigma(\textrm{Im}(z))&=&-\textrm{Im}(z) \ , \nn\\
\sigma(\textrm{Im}(\Omega_2))&=&-\textrm{Im}(\Omega_2) \ , \nn\\
j_{||}&=&\gamma\ \textrm{Re}(\Omega_2)_{||} \ , \nn\\
j_{\bot}&=&-\frac{1}{\gamma}\ \textrm{Re}(\Omega_2)_{\bot} \ .
\eea
In this form, the projection conditions are now much more tractable.

\subsection{The pure spinors and the projection basis}\label{purespindiel}

In this subsection, we will rewrite the pure spinors in terms of the variables of the projection basis.
But before going back to the pure spinors, let us first give some useful relations
(they are nothing else but a rewriting of the two last projection conditions
given in (\ref{projb1})):
\bea
\label{useful}
\textrm{IIA}&:& \quad \! k_{||}j_{||}+k_{\bot}\textrm{Re}(\Omega_2)_{||}=0 \ , \qquad -k_{\bot}j_{\bot}+k_{||}\textrm{Re}(\Omega_2)_{\bot}=0 \ ,  \nn \\
\textrm{IIB}&:&  -k_{\bot}j_{||}+k_{||}\textrm{Re}(\Omega_2)_{||}=0 \ , \quad \! \qquad k_{||}j_{\bot}+k_{\bot}\textrm{Re}(\Omega_2)_{\bot}=0 \ .
\eea
These allow to write the following relations valid for both theories:
\bea
\label{rel}
&& -\sin(\phi)j+\cos(\phi)\textrm{Re}(\Omega_2)=\frac{1}{\cos(\phi)}\textrm{Re}(\Omega_2)_{\bot}=-\frac{1}{\sin(\phi)}j_{\bot} \ , \nn \\
&& \quad \! \cos(\phi)j+\sin(\phi)\textrm{Re}(\Omega_2)=\frac{1}{\sin(\phi)}\textrm{Re}(\Omega_2)_{||}=\frac{1}{\cos(\phi)}j_{||} \ .
\eea
These last relations (\ref{rel}) can also be found by using the definitions of $j_{||}$, $j_{\bot}$, $\textrm{Re}(\Omega_2)_{||}$, and $\textrm{Re}(\Omega_2)_{\bot}$. One can notice in the previous relation a rotation. We will come back to it soon.\\

We can now rewrite the pure spinors in (\ref{type}) using the projection basis
and the relations (\ref{rel}). The result is very simple:
\bea
\textrm{IIA}&:& \Phi_+ = \frac{|a|^2}{8}e^{-i\theta}k_{||}\ e^{\frac{1}{2} z\w \overline{z}-\frac{i}{k_{||}k_{\bot}}\textrm{Re}(\Omega_2)_{\bot}+\frac{k_{\bot}}{k_{||}}\textrm{Im}(\Omega_2)} \ , \nn \\
& & \Phi_- = -\frac{|a|^2}{8}k_{\bot}\ z\w e^{\frac{i}{k_{||}k_{\bot}}\textrm{Re}(\Omega_2)_{||}-\frac{k_{||}}{k_{\bot}}\textrm{Im}(\Omega_2)} \ , \label{typeA} \\
\textrm{IIB}&:& \Phi_+ = \frac{|a|^2}{8}e^{-i\theta}k_{||}\ e^{\frac{1}{2} z\w \overline{z}-\frac{i}{k_{||}k_{\bot}}\textrm{Re}(\Omega_2)_{||}+\frac{k_{\bot}}{k_{||}}\textrm{Im}(\Omega_2)} \ , \nn \\
& & \Phi_- = -\frac{|a|^2}{8}k_{\bot}\ z\w e^{\frac{i}{k_{||}k_{\bot}}\textrm{Re}(\Omega_2)_{\bot}-\frac{k_{||}}{k_{\bot}}\textrm{Im}(\Omega_2)} \ . \label{typeB}
\eea

Recently, an alternative parametrization of the internal supersymmetry parameters,
and consequently of the pure spinors, was given in
\cite{MPZ} and further discussed in \cite{Al}
\bea
\label{dielans}
\eta_{+}^1 &=& a\left (\cos(\Psi)\eta_{+D}-\sin(\Psi) \frac{z\eta_{-D}}{2} \right ) \ , \nn\\
\eta_{+}^2 &=& ae^{i\theta}\left (\cos(\Psi)\eta_{+D}+\sin(\Psi) \frac{z\eta_{-D}}{2}
\right ) \ ,
\eea
where we still have $\theta$ as the difference of phase between $\eta_{+}^2$ and $\eta_{+}^1$, $a$ and $z$ are the same as before, $||\eta_{+D}||=1$ and $\Psi$ is an angle such as $0\leq \Psi \leq \frac{\pi}{4}$. This different choice was proposed in order to study deformations of four-dimensional $\mathcal{N}=4$ Super Yang-Mills in the context of AdS/CFT. Typically those deformations should describe the near horizon geometry of some sort of dielectric branes, hence the name dielectric for the  spinor $\eta_{+D}$. Note that $\eta_{+D}$ is nothing else but (once the phases of the two spinors are equalled) the mean spinor between $\eta_{+}^1$ and $\eta_{+}^2$, i.e. somehow their bisector: $\eta_{+D}=\frac{1}{2a\ \cos(\Psi)}(\eta_{+}^1+e^{-i\theta}\eta_{+}^2)$. We have the corresponding picture:
\begin{figure}[H]
\begin{center}
\psfrag{Psi}[][]{$\Psi$}
\psfrag{eta1}[][]{$\frac{\eta_+^1}{a}$}
\psfrag{eta2}[r][]{$\frac{\eta_+^2}{a}$}
\psfrag{etaD}[][b]{$\eta_{+D}$}
\psfrag{etaDo+}[][]{$\frac{z.\eta_{-D}}{2}$}
\includegraphics{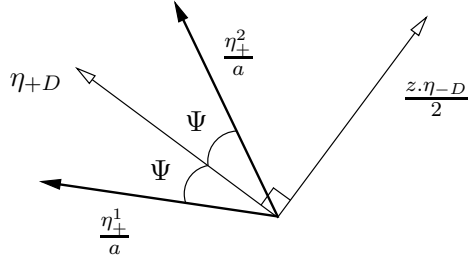}
\caption{The different spinors and angles (with $\theta=0$)}
\end{center}
\end{figure}
One can relate the dielectric ansatz to the previous one, (\ref{defk}), with
\beq
k_{||}=\cos(\phi)= \cos(2\Psi),\ k_{\bot}=\sin(\phi) = \sin(2\Psi)\label{phipsi} \ ,
\eeq
\beq
\eta_{+D}=\cos(\frac{\phi}{2}) \eta_{+} + \sin(\frac{\phi}{2}) \frac{z\eta_{-}}{2} \label{eta+D} \ .
\eeq
Working with $\eta_{+D}$ and $\frac{z \eta_{-D}}{2}$ instead of $\eta_+$ and $\frac{z \eta_-}{2}$ means working with a new $SU(2)$ structure. And this new $SU(2)$ structure is clearly obtained by a rotation from the previous one, as one can also get by computing the relations between the $SU(2)$ structure two-forms:
\bea
\label{dielecform}
j_D &=& k_{||}j+k_{\bot} \textrm{Re}(\Omega_2) \ , \nn \\
\textrm{Re}(\omega_D) &=& -k_{\bot} j+k_{||}\textrm{Re}(\Omega_2) \ , \nn \\
\textrm{Im}(\omega_D) &=& \textrm{Im}(\Omega_2) \ .
\eea

If one computes the pure spinors from (\ref{dielans}) \cite{MPZ, Al}, one gets
the dielectric pure spinors\footnote{The computation is the same as using (\ref{type}) and introducing the dielectric $SU(2)$ structure variables via (\ref{dielecform}).}
\bea
\label{su2purespinors}
\Phi_+ &=& \frac{|a|^2}{8}e^{-i\theta}k_{||}\ e^{\frac{1}{2} z\w \overline{z}-\frac{i}{k_{||}}j_D+\frac{k_{\bot}}{k_{||}}\textrm{Im}(\omega_D)} \ , \nn \\
\Phi_- &=& -\frac{|a|^2}{8}k_{\bot}\ z\w e^{\frac{i}{k_{\bot}}\textrm{Re}(\omega_D)-\frac{k_{||}}{k_{\bot}}\textrm{Im}(\omega_D)} \ .
\eea

Comparing the definitions of the two-forms (\ref{rel}) and (\ref{dielecform}), or the expressions for the pure spinors, (\ref{typeA}), (\ref{typeB}) and (\ref{su2purespinors}), we see that (for IIA/IIB)
\bea
j_D &=& \frac{1}{k_{\bot}}\textrm{Re}(\Omega_2)_{\bot/||} \ , \nn \\
\textrm{Re}(\omega_D) &=& \frac{1}{k_{||}}\textrm{Re}(\Omega_2)_{||/\bot} \ , \nn\\
\textrm{Im}(\omega_D) &=& \textrm{Im}(\Omega_2)\label{link} \ .
\eea
Thus the dielectric $SU(2)$ structure variables are nothing but the eigenbasis of the orientifold projection (modulo a rescaling) ! Actually, this can be easily understood from the transformation properties of $\eta_{+D}$ under the orientifold projection\footnote{To get them, we recall that we have $e^{i\theta}=\pm1$ for an $O5$, and one has to use (\ref{projeta1}) and (\ref{projeta2}).}
\bea
O6&:& \sigma(\eta_{\pm D})=\eta_{\mp D} \ , \nn \\
O5&:& \sigma(\eta_{\pm D})=e^{i\theta}\eta_{\pm D} \ .
\eea
Then the $SU(2)$ bilinears constructed from it will get at most a phase and a conjugation
when being applied $\sigma$, hence the three real two-forms $j_D$, $\textrm{Re}(\omega_D)$ and $\textrm{Im}(\omega_D)$ are in the projection eigenbasis, as given by (\ref{link}). Note that these relations between those variables is a way to understand the rotation that gives the projection basis, as mentioned after (\ref{rel}).\\

Beside providing a tractable basis to solve the orientifold projection conditions, the dielectric
variables/projection basis lead to simpler expressions of the pure spinors and so much simpler SUSY conditions (see next subsection). Hence this $SU(2)$ structure is a much better choice to solve our problem, and we will express the equations to be solved in terms of these variables. For instance, in next subsection, we rewrite the SUSY conditions in terms of the projection basis. And in appendix \ref{projbasis}, we rewrite a set of $SU(2)$ structure conditions (implying the compatibility conditions, see appendix \ref{proofcom}) in terms of the projection basis variables too.

\subsection{SUSY equations in the projection basis}\label{SUSYsec}

In appendix \ref{SUSYap}, we give the SUSY equations (\ref{susy1}), (\ref{susy2}), and (\ref{susy3}), expanded in forms for general expressions of the pure spinors. Here we consider
a simplified version of those equations where beside the usual fixing of the parameters leading to (\ref{purespin}), we choose $|a|^2=e^A$, and go to the large volume limit, i.e. $A=0$ and $e^\phi=g_s$ constant. This is indeed the regime in which we will look for solutions in the next section.
The only remaining freedom is $\theta$ that we do not really need to fix. Moreover, we choose to look only for intermediate $SU(2)$ structure, i.e. with $k_{||}\neq 0$ and $k_{\bot}\neq 0$ and constant.
Taking the coefficients constant is important because it simplifies drastically the search
for solutions (the SUSY conditions are much simpler), but forbids to get genuinely dynamical $SU(2)$ structure vacua.\\

Using the projection basis variables and some further simplifications explained in appendices \ref{SUSYap} and \ref{More}, the supersymmetry equations finally become
\bea
\textrm{IIA}&:& g_s*F_4=-k_{\bot}d(\textrm{Im}(z)) \nn\\
& & k_{||}H=k_{\bot}d(\textrm{Im}(\Omega_2)) \nn\\
& & g_s*F_2=-k_{||}d(\textrm{Im}(z))\w \textrm{Im}(\Omega_2)+\frac{1}{k_{||}}d(\textrm{Re}(\Omega_2)_{||})\w \textrm{Re}(z)-\frac{1}{k_{\bot}}H\w \textrm{Im}(z) \nn\\
& & g_s*F_0=\frac{1}{2}k_{\bot}d(\textrm{Im}(z))\w \textrm{Im}(\Omega_2)^2+\frac{1}{k_{||}} H\w \textrm{Re}(z)\w \textrm{Re}(\Omega_2)_{||} \nn\\
\nn\\
\nn\\
& & d(\textrm{Re}(z))=0 \nn\\
& & d(\textrm{Re}(\Omega_2)_{\bot})=k_{||}k_{\bot} \textrm{Re}(z)\w d(\textrm{Im}(z)) \nn\\
& & H\w \textrm{Re}(z)=-\frac{k_{\bot}}{k_{||}}d(\textrm{Im}(z)\w \textrm{Re}(\Omega_2)_{||}) \ , \label{SUSYA}\\
\nn\\
\nn\\
\textrm{IIB}&:& k_{||}H=k_{\bot}d(\textrm{Im}(\Omega_2)) \nn\\
& & k_{\bot}e^{i\theta}g_s*F_3=d(\textrm{Re}(\Omega_2)_{||}) \nn\\
& & k_{\bot}e^{i\theta}g_s*F_1=H\w \textrm{Re}(\Omega_2)_{||} \nn\\
\nn\\
\nn\\
& & d(\textrm{Re}(z))=0 \nn\\
& & d(\textrm{Im}(z))=0 \nn\\
& & \textrm{Re}(z)\w H=-\frac{k_{\bot}}{k_{||}}\textrm{Im}(z)\w d(\textrm{Re}(\Omega_2)_{\bot}) \nn\\
& & \textrm{Im}(z)\w H=\frac{k_{\bot}}{k_{||}}\textrm{Re}(z)\w d(\textrm{Re}(\Omega_2)_{\bot}) \nn\\
& & \textrm{Re}(z)\w \textrm{Im}(z) \w d(\textrm{Re}(\Omega_2)_{||})=- H\w \textrm{Im}(\Omega_2) \ .\label{SUSYB}
\eea

\section{Solutions}\label{secsol}

\subsection{Set-up, method, and discussion}\label{Solconv}

In \cite{Scan}, examples of four-dimensional Minkowski supersymmetric flux vacua, with a Generalized Calabi-Yau as internal manifold, were found: they correspond to nilmanifolds and solvmanifolds\footnote{Nil/solvmanifolds, also known as twisted tori, can be seen as iterated fibrations of tori over other tori. They are parallelizable manifolds, namely they admit a basis of real globally defined one-forms, which we will note $e^i,\ i=1..6$. These manifolds are group manifolds, and can be defined by their ``algebra''. We will use for it the following notation: $(0,0,0,12,23,14-35)$, for instance, means $de^1=de^2=de^3=0$, $de^4=e^1\w e^2$, etc. with $d$ the exterior derivative. For more details on these manifolds, see for instance \cite{Scan}.} with non trivial fluxes. As already mentioned in introduction, the analysis of \cite{Scan} did not take into account the possibility of an intermediate $SU(2)$ structure on the internal manifold. Some examples of solutions with such a structure were found in \cite{KT} via T-dualities from a warped $T^6$ with an $O3$. In this paper, we extend the analysis of \cite{Scan} and find, among nil/solvmanifolds, new vacua with intermediate $SU(2)$ structure that cannot be T-dualized back to a warped $T^6$ with an $O3$.\\

Before describing our new solutions, we briefly sketch the method we followed. We first choose a nil/solvmanifold among the list given in \cite{Scan}, we specify the theory (IIA/IIB) and the internal directions of an O-plane. In this paper we will only consider O5- and O6-planes (see beginning of section \ref{secproj}). The orientifold projection should be compatible with the manifold algebra (see \cite{Scan} for the complete list of the allowed orientifolds for each manifold). Then, one has to find a pair of compatible pure spinors on the internal manifold. The general form of the pure spinors is given in (\ref{purespin}) where, in order to have an intermediate $SU(2)$ structure, we take $k_{||}\neq 0$ and $k_{\bot}\neq 0$, and constant. The other coefficients in the solutions will also be taken constant. Moreover, we choose $|a|^2=e^A$, and go to the large volume limit, i.e. where $A=0$ and $e^\phi=g_s$ is constant\footnote{In \cite{Scan}, they give a method to localize the solutions obtained by reintroducing afterwards the warp factor. But these techniques only work for solutions with one source, while we will obtain solutions with two sources. So we will not try to get solutions in another regime than in the large volume limit, with smeared sources and constant coefficients in the solutions. As discussed for the SUSY equations, this forbids to obtain genuinely dynamical $SU(2)$ structure solutions.}. We will use the set of new variables, the projection basis, which corresponds to the appropriate $SU(2)$ structure in this problem, since many equations written in these variables get simplified (see section \ref{secproj}). We then solve the projection conditions (\ref{projb1}) so that these pure spinors are compatible with the O-plane, and then the $SU(2)$ structure conditions (\ref{com3}) to (\ref{com7}), and (\ref{comb1}), getting automatically that the pure spinors are compatible (see appendix \ref{proofcom}).\\

This pair must satisfy the SUSY conditions, implying that one of them is closed and the
manifold is thus a GCY. Still using the projection basis, we then solve the SUSY equations (\ref{SUSYA}) or (\ref{SUSYB}). For every solution, we can then introduce a local basis of complex one-forms $(z^1,z^2,z,\overline{z}^1,\overline{z}^2,\overline{z})$, where we identify one of them with the holomorphic one-form $z$ of the $SU(2)$ structure, and write the real and the holomorphic two-forms of the $SU(2)$ structure as
\beq
\Omega_2= z^1\w z^2 \qquad j=\frac{i}{2} ( t_1 z^1\w \overline{z}^1 + t_2 z^2\w \overline{z}^2 + b z^1\w \overline{z}^2 - \overline{b} \overline{z}^1 \w z^2 ) \ ,\label{factorO2}
\eeq
with $b=b_r+ib_i$ and $t_1$, $t_2$, $b_r$, $b_i$ real\footnote{Note that the choice of this basis is not unique. This freedom will appear in particular in the limits (subsection \ref{Continuation}).}. We will give our solutions in the previous form\footnote{Note that the metric we will then compute from it will be block diagonal, so the left $SU(2)$ structure conditions, namely the contractions with $z$ and $\overline{z}$, are clearly satisfied by these expressions.}.\\

With the almost complex structure (see footnote \ref{holo}) defined trivially in the local complex basis $(z^1,z^2,z,\overline{z}^1,\overline{z}^2,\overline{z})$ by $J^{\ \ \lambda}_\mu =i\delta^{\ \ \lambda}_\mu,\ J^{\ \ \overline{\lambda}}_{\overline{\mu}}=-i\delta^{\ \ \overline{\lambda}}_{\overline{\mu}}$ ((anti)holomorphic indices), and the K\"ahler form defined as in (\ref{SU2}), one can then compute the hermitian metric:
\beq
g_{\mu\overline{\nu}}=-J^{\ \ \lambda}_\mu J_{\lambda\overline{\nu}} \qquad g_{\overline{\mu}\nu}=-J_{\overline{\mu}}^{\ \ \overline{\lambda}}J_{\overline{\lambda}\nu} \ .
\eeq
In this local complex basis, we obtain generically\footnote{Note that we give here the coefficients of the metric tensor: they are symmetric, but do not have to be real, since only the tensor has to be real. To get the metric in the real basis $(e^i,\ i=1..6)$, one has to perform a change of basis.}
\beq
g=\frac{1}{2} \left (\begin{array}{cccccc} 0 & 0 & 0 & t_1 & b & 0 \\ 0 & 0 & 0 & \overline{b} & t_2 & 0 \\ 0 & 0 & 0 & 0 & 0 & 1 \\ t_1 & \overline{b} & 0 & 0 & 0 & 0 \\ b & t_2 & 0 & 0 & 0 & 0 \\ 0 & 0 & 1 & 0 & 0 & 0 \end{array}\right )
\eeq
To check its definite-positiveness, one has to verify that for any $\mu$, $g^{\mu \overline{\mu}}>0$ (coefficients of the inverse metric), which is equivalent to $\frac{t_i}{(t_1t_2-|b|^2)}>0$, $i=1,\ 2$. Actually, the $SU(2)$ structure condition (\ref{SU2vol}), that the solutions verify, gives that $t_1t_2-|b|^2=1$. Hence the
definite-positiveness of the metric becomes equivalent to $t_1>0$ and $t_2>0$.\\

The final step is to compute the RR fluxes, defined by the last SUSY equation (\ref{susy3}), and to check whether they solve the Bianchi identities. Note that the metric is needed to compute the RR fluxes, because of the Hodge star. We compute the BI, and then we can determine the sources and their charges (see (\ref{BI2})). Since the sources are smeared, the BI will give us directly the directions of the co-volume $V^i$ of the cycles wrapped by the sources. We only have to compute the correct normalization of these co-volumes. To do so, we use, as done in \cite{Scan}, the following identity, motivated in appendix \ref{cali}, and built from the calibration of the sources \cite{Calib, KT}:
\beq \left\langle V^i, e^{3A-\phi} \textrm{Im}(\Phi_2) \right\rangle = \frac{1}{8g_s} \ V \ ,\label{normVi}\eeq
where $V$ is the internal volume form, defined the following way (see (\ref{compcond})):
\beq \left\langle \Phi_\pm, \overline{\Phi}_\pm \right\rangle=-\frac{i}{8} ||\eta_+^1||^2 ||\eta_\pm^2||^2 V \ , \label{volume}\eeq
and we have $\int_{M_6} V >0$. Note that this normalization condition is not exactly the same as in \cite{Scan}\footnote{In \cite{Scan}, they did not have the $\frac{1}{8g_s}$ factor, that we explain in appendix \ref{cali}.}. Once we have identified $V^i$, we deduce the source charge\footnote{Note that using this condition and our conventions for the Hodge star, it can be shown as in \cite{Scan} that $\sum_{i} Q_i < 0$, and so recover the need for orientifolds as sources, because of their negative charge.}. If it is negative, we deduce we have an orientifold, and we verify that the manifold and our solution are compatible with its projection.\\

We would like to stress that our search, on nil/solvmanifolds, for solutions with intermediate $SU(2)$ structure is not meant to be exhaustive, our interest being to verify the possibility of having solutions of this kind that are not obtainable via T-duality. We decided to look at the manifolds for which non T-dual solutions with $SU(3)$ or static $SU(2)$ structure were found in \cite{Scan} (the nilmanifold (0,0,0,12,23,14-35), noted $n\ 3.14$, and the solvmanifold (25,-15, $\alpha$ 45, - $\alpha$ 35, 0,0), noted $s\ 2.5$) with the intuitive hope that some intermediate $SU(2)$ structure might be found on them, which might give back their solutions in the limits $k_{\bot/||}\rightarrow 0$. We indeed find three new solutions which we will describe in the next subsection. In  subsection \ref{Continuation}, we discuss their possible limits to the solutions of \cite{Scan}. Note that these
solutions cannot be T-dualized back to a warped $T^6$ with an $O3$ for the same reason as in \cite{Scan}: one should T-dualize back along the internal directions chosen for the $O5$ or the $O6$. But one can see from the algebras of the manifolds that there is no isometry in these directions (an isometry direction should not appear in the algebra). In subsection \ref{other}, we will discuss the possibility of finding some other solutions.\\

Finally, let us say a word about the directions chosen for the orientifolds in our solutions. In \cite{Scan} they give, for each manifold to be considered, the orientifolds compatible with the algebra (i.e. the involution $\sigma$ due to the O-plane must commute with the algebra). On the two manifolds we are going to consider at first, here are the possible directions for the $O5$ and the $O6$:
\begin{table}[H]
\begin{center}
\begin{tabular}{c|c|c} \textrm{Manifold} & $O5$ & $O6$ \\ \hline n\ 3.14 & 13, 15, 26, 34, 45 & \textrm{none} \\ \hline s\ 2.5 & 13, 14, 23, 24, 56 & 125, 136, 146, 236, 246, 345
\end{tabular}\caption{Directions of the possible $O5$ and $O6$ on the manifolds considered}\label{Dir}
\end{center}
\end{table}
Among these possibilities, we are going to look for solutions only for one set of directions on each manifold. So one could ask about the other directions. As explained in subsection \ref{other} and in appendix \ref{reducset}, one can actually consider the symmetries of the algebra to relate several possible O-planes. Furthermore, if one looks for solutions with several (not completely overlapping) O-planes, one can prove, as we do in appendix \ref{reducset}, that it is enough to look for solutions with the sources in the directions we are going to choose.

\subsection{Intermediate $SU(2)$ solutions}

We are now going to give the solutions found, with detailed steps for the first solution, and then quicker for the two others.

\subsubsection{First solution}

We look for IIB solutions on the nilmanifold $n\ 3.14$ which has the following algebra: $(0,0,0,12,23,14-35)$, with an $O5$ in the $45$ directions. We find the general solutions to the list of constraints (\ref{projb1}), (\ref{com3}), (\ref{com4}), (\ref{com5}), (\ref{com6}), (\ref{com7}), (\ref{comb1}) and (\ref{SUSYB}). The solutions depend on the following real (constant) parameters: $b_{12},b_{23},b_{26},b_{24},b_{46},c_{24},c_{46},f_1,f_3,k_{||}$ ($k_{\bot}$ can be replaced everywhere by $\sqrt{1-k_{||}^2}$). These parameters have to satisfy certain conditions so that the solution is genuinely one: $b_{26}$ and $b_{24}c_{46}-c_{24}b_{46}$ have to be non-zero, $f_1$ or $f_3$ has to be non-zero, $k_{||}$ has to be nor $0$ neither $1$. As explained, after finding the solutions, we expressed them as in (\ref{factorO2}) with for the first solution:
$$ z = (f_1+if_3)e^1+(f_3-if_1)e^3\ , $$
$$ z^1 = b_{12}e^1-b_{23}e^3-(b_{24}+ic_{24})e^4+ \left (-i(b_{24}+ic_{24}) +\frac{k_{\bot}^2 (b_{46}-ic_{46}) (b_{24}c_{46}-c_{24}b_{46})}{b_{46}^2+k_{||}^2 c_{46}^2} \right) e^5 -b_{26} e^6\ , $$
$$ z^2 = e^2+\frac{(b_{46}+ic_{46})}{b_{26}}e^4+i\frac{(b_{46}+ic_{46})}{b_{26}}e^5\ , $$
$$ b_r = -\frac{b_{24}b_{46}+c_{46}c_{24}k_{||}^2}{(b_{24}c_{46}-c_{24}b_{46})k_{||}k_{\bot}}\ , \qquad b_i = -\frac{k_{||}}{k_{\bot}}\ , $$
\beq t_1 = -\frac{b_{46}^2+c_{46}^2k_{||}^2}{b_{26}(b_{24}c_{46}-c_{24}b_{46})k_{||}k_{\bot}}\ , \qquad t_2 = -\frac{(b_{24}^2+c_{24}^2k_{||}^2)b_{26}}{(b_{24}c_{46}-c_{24}b_{46})k_{||}k_{\bot}} \ .\eeq

There is a second solution which is obtained from the first one by conjugating $z$ and doing $e^5 \rightarrow -e^5$. The conditions on the coefficients for this second solution are the same.\\

As explained, the definite-positiveness of the metric is given by
\beq b_{26}(b_{24}c_{46}-c_{24}b_{46})<0 \ . \label{defpos}\eeq

For the general solution given before, we have
\beq
H=\frac{k_{\bot}}{k_{||}} \left (\frac{c_{46} b_{23}-b_{46} b_{12}}{b_{26}} e^1\w e^2 \w e^3 + c_{46} (e^1 \w e^2 \w e^6 -e^3 \w e^4 \w e^5) - b_{46} ( e^1 \w e^4 \w e^5 + e^2\w e^3\w e^6) \right ) \ .
\eeq

The general metric in the real basis can be computed with the method described previously, and its determinant $|g|$ is:
\beq \frac{(f_1^2+f_3^2)^2 (-b_{46} c_{24}+c_{46} b_{24})^2 (c_{46}^2+b_{46}^2)^2}{(b_{46}^2+k_{||}^2 c_{46}^2)^2} \label{detsol1}\eeq
(clearly non-zero). The general expression of the metric is actually quite complicated because there are many parameters, so we will not give it here. Furthermore it is difficult to compute properly its eigenvalues, and then, to use them to compute the Bianchi identities. So let us go to a simpler case, in order to show that there is at least one solution. To do so, we can make the following allowed choice of the solution's parameters:
\beq b_{12}=b_{23}=b_{46}=c_{24}=0 \ . \label{simpsol1} \eeq
This choice is interesting because then, the metric becomes diagonal (in the $e^i$ basis !): its coefficients are given by\footnote{Note that our convention $||z||^2=\overline{z}^\mu z_\mu=2$ is already implemented in the metric, by its construction from the K\"ahler form in which this norm appears. One can verify this point by computing this norm using either the hermitian or the real basis metric. Then, $f_1^2+f_3^2$ has nothing to do with this norm, but is only the measure related to the metric coefficients, in the real basis.}:
\beq
g=\left (\begin{array}{cccccc} f_1^2+f_3^2 & 0 & 0 & 0 & 0 & 0 \\ 0 & -\frac{b_{24}b_{26}}{c_{46}k_{||}k_{\bot}} & 0 & 0 & 0 & 0 \\ 0 & 0 & f_1^2+f_3^2 & 0 & 0 & 0 \\ 0 & 0 & 0 & -\frac{b_{24}c_{46}k_{\bot}}{b_{26}k_{||}} & 0 & 0 \\ 0 & 0 & 0 & 0 & -\frac{b_{24}c_{46}k_{\bot}}{k_{||}^3b_{26}} & 0 \\ 0 & 0 & 0 & 0 & 0 & -\frac{k_{||}c_{46}b_{26}}{k_{\bot}b_{24}} \end{array}\right )
\eeq
Notice that with these eigenvalues, we can recheck the definite-positiveness of the metric, and we get the same condition as the one found before (\ref{defpos}) with the hermitian metric: the eigenvalues are strictly positive if and only if
\beq b_{26}b_{24}c_{46}<0 \ .\eeq

To get the Bianchi identities, we first have to be able to perform a (six-dimensional) Hodge star $*$ to get the RR fluxes (see the definitions of the fluxes in the SUSY conditions (\ref{SUSYB})), that is where the metric is used (see appendix \ref{Convdiff} for the conventions on the Hodge star). When we have a RR flux, we can then compute the Bianchi identity, and then we identify the sources obtained (see subsection \ref{Solconv}) and see whether the O-planes are compatible with the manifold and the solution. Here, we get the following fluxes:
$$H=\frac{k_{\bot}c_{46}}{k_{||}}\left (-e^3\w e^4\w e^5 + e^1\w e^2\w e^6 \right ) \ ,$$
$$F_3=\frac{e^{-i\theta} |c_{46}| k_{||}^2}{g_s |b_{24}|}\left ( -\frac{k_{\bot}b_{24}c_{46}}{k_{||}^2b_{26}}\ (e^3\w e^4\w e^6 +\frac{1}{k_{||}^2}e^1\w e^5\w e^6) +\frac{b_{26}}{k_{\bot}}\ (-\frac{1}{k_{||}^2} e^3\w e^5\w e^6 + e^1\w e^4\w e^6) \right ) \ ,$$
\beq F_1=\frac{e^{-i\theta}c_{46} k_{||}}{g_s |c_{46}b_{24}|} \left ( -b_{26} e^1+\frac{b_{24} c_{46} k_{\bot}^2}{k_{||}^2 b_{26}} e^3 \right )\ .\eeq

We then compute the Bianchi identities:
$$d(F_1)=0 \ ,\ H\w F_3=0 \ ,$$
\beq d(F_3)-H\w F_1= \frac{2 e^{-i\theta} |c_{46}|}{g_s |b_{24}| k_{||}^2 k_{\bot}} \left (\frac{k_{\bot}^2 b_{24} c_{46}}{b_{26}}\ e^1 \w e^2 \w e^3 \w e^6+b_{26} k_{||}^2\ e^1\w e^3\w e^4\w e^5 \right) \ . \eeq
We see that there is no source for $F_1$ (neither for $F_5$), which is somehow expected as we did not put any. We see that $F_3$ has two sources, one along the directions $45$ and the other along $26$. As explained in subsection \ref{Solconv}, to determine their charges, we first need to compute their co-volumes. To do so, we first compute $V$. From (\ref{volume}), using (\ref{purespin}) for the pure spinors, and then the form (\ref{factorO2}) of the solutions, we get:
\beq
V=\frac{1}{8i} z\w \overline{z} \w \Omega_2 \w \overline{\Omega}_2 =- \textrm{Re}(z^1) \w \textrm{Im}(z^1) \w \textrm{Re}(z^2) \w \textrm{Im}(z^2) \w \textrm{Re}(z) \w \textrm{Im}(z) \ .
\eeq
Note that going to the real basis given by the $e^i$, by replacing the $z^i$ for each solution found, one generically gets:
\beq V=C\ e^1 \w e^2 \w e^3 \w e^4 \w e^5 \w e^6 \ . \eeq
We chose the orientability conventions $\epsilon_{123456}=1$ (see appendix \ref{Convdiff}), so $C$ is clearly related to $\sqrt{|g|}$, but also to the determinant\footnote{Note that checking $|g|\neq 0$ then verifies that the $z^i$ chosen form indeed a basis. In general, we actually already have it guaranteed because of (\ref{comb1}).} of the matrix allowing to go from the $z^i$ to the $e^i$. So we get $C >0$, an important point to determine the sign of the charges. Having computed $V$, one can determine the $V^i$ precisely using the relation (\ref{normVi}): we know already that $V^i$ is along the transverse directions of the source, and (\ref{normVi}) gives the normalization factor. Note one can rewrite (\ref{normVi}), using (\ref{purespin}) in the large volume limit, as
\bea
\textrm{IIA}&:& V^i\w \left(\textrm{Re}(z) \w \textrm{Re}(\Omega_2)_{||} - k_{||}^2\ \textrm{Im}(z) \w \textrm{Im}(\Omega_2)\right)= k_{||}\ V \ ,\nn\\
\textrm{IIB}&:& V^i\w \left(\textrm{Re}(\Omega_2)_{||} +k_{\bot}k_{||}\ \textrm{Re}(z)\w \textrm{Im}(z)\right)= k_{\bot}e^{i\theta}\ V \ . \label{Vi}
\eea
Finally, for the first solution, we can rewrite the BI as:
\beq V^1=-\frac{k_{\bot}e^{i\theta}(f_1^2+f_3^2)b_{26}}{k_{\bot}^2}\ e^1\w e^2 \w e^3 \w e^6,\ V^2=-\frac{k_{\bot}e^{i\theta}(f_1^2+f_3^2)b_{24}c_{46}}{k_{||}^2 b_{26}}\ e^1\w e^3 \w e^4 \w e^5 \ ,\eeq
\beq d(F_3)-H\w F_1= -\frac{2|c_{46}|}{g_s |b_{24}| k_{||}^2 C} \left (\frac{(k_{\bot} b_{24} c_{46})^2}{b_{26}^2 k_{||}^2}\ V^1+ \frac{(k_{||} b_{26})^2}{k_{\bot}^2}\ V^2 \right),\ \textrm{with}\  C=\frac{(f_1^2+f_3^2)b_{24}c_{46}}{k_{||}^2} >0 \ . \eeq
So one can read directly the charges (see (\ref{BI2})) and see that $Q_1 <0$, $Q_2 <0$, hence we have two O-plane sources. Both are compatible with the manifold. Note that it is interesting to see this second source appearing while we only imposed the first one.\\

With the choice made for the parameters, the solution is  (we do not display $j$ since it is deduced easily from $\Omega_2$ with the projection conditions):
\bea
\textrm{Re}(\Omega_2)_{||} &=& -\frac{b_{24}c_{46} k_{\bot}^2}{k_{||}^2 b_{26}} e^4\w e^5+b_{26} e^2\w e^6 \ ,\nn\\
\textrm{Re}(\Omega_2)_{\bot} &=& b_{24} e^2\w e^4-c_{46} e^5\w e^6 \ ,\nn\\
\textrm{Im}(\Omega_2) &=& \frac{b_{24}}{k_{||}^2} e^2\w e^5+c_{46} e^4\w e^6 \ .
\eea
It is clear from this formulation that what is parallel or orthogonal under $\sigma_{45}$ is also under $\sigma_{26}$. The same goes for $z$ which only has components along $e^1$ and $e^3$. So the solution is clearly compatible with the projections of both sources.\\

Note that we will not find any T-dual solution to this first solution, while the two next solutions are T-duals to one another. This can be understood from table \ref{Dir} since no $O6$ is compatible with $n\ 3.14$.

\subsubsection{Second solution}

We proceed in the same way as for the first solution. We look for IIB solutions on the solvmanifold $s\ 2.5$ which has the following algebra: $(25,-15,\alpha 45,-\alpha 35, 0, 0)$, $\alpha \ \in \ \mathbb{Z}$, with an $O5$ in the $13$ directions. The general solution to the usual list of constraints depends on the following real (constant) parameters: $b_{25},b_{45},b_{24},b_{12},b_{23},c_{12},c_{23},f_5,f_6,g_5,g_6,k_{||}$, and of course $\alpha$. These parameters have to satisfy certain conditions so that the solution is genuinely one: $b_{24}$, $f_5g_6-f_6g_5$ and $c_{23}b_{12}-c_{12}b_{23}$ have to be non-zero, $k_{||}$ has to be neither $0$ nor $1$, and $\alpha$ has to be $\pm1$. The solution is expressed in the usual manner with the following $z^i$:
$$z=(f_5+i g_5)e^5+(f_6+i g_6)e^6 \ ,$$
$$z^1=(b_{12}+i c_{12}) e^1- (b_{23}+i c_{23}) e^3 -b_{24} e^4 -b_{25} e^5 \ ,$$
$$z^2=e^2 + \frac{\alpha k_{\bot}^2 (c_{23}b_{12}-c_{12}b_{23})^2}{(b_{12}^2 +c_{12}^2 k_{||}^2)(b_{12}+i c_{12})b_{24}}e^3-\frac{\alpha (b_{23}+i c_{23})}{(b_{12}+i c_{12})}e^4 - \frac{1}{b_{24}}\left ( b_{45} + \frac{\alpha b_{25}(b_{23}+i c_{23})}{(b_{12}+i c_{12})} \right ) e^5 \ ,$$
$$b_r=\frac{k_{\bot} b_{12} c_{12}}{k_{||}(b_{12}^2+c_{12}^2)}\ , \qquad b_i=\frac{k_{\bot} b_{12}^2 }{k_{||}(b_{12}^2+c_{12}^2)} \ ,$$
\beq t_1=-\frac{\alpha k_{\bot}(c_{23}b_{12}-c_{12}b_{23})}{k_{||}b_{24}(b_{12}^2+c_{12}^2)}\ , \qquad
t_2=-\frac{b_{24}(b_{12}^2+c_{12}^2 k_{||}^2)}{\alpha k_{\bot}k_{||}(c_{23}b_{12}-c_{12}b_{23})}\ . \eeq

The definite-positiveness of the metric is given by
\beq \alpha b_{24} (c_{23}b_{12}-c_{12}b_{23})<0 \ .\label{defpos2}\eeq

For the general solution given before, and given that $\alpha^2=1$, we have $H=0$, and deduce
\beq F_1=0 \ .\eeq

The only remaining flux is then $F_3$. As for the first solution, the general metric is quite complicated, and it is difficult to compute its eigenvalues, so we will go to a simpler case. We just mention here its determinant, once again clearly non-zero:
\beq \frac{(b_{23}c_{12}-c_{23}b_{12})^4(f_5g_6-g_5f_6)^2}{(c_{12}^2 k_{||}^2+b_{12}^2)^2}\ .\label{detsol2}\eeq

To simplify the metric, we first choose $b_{25}=b_{45}=0$. Then to get a diagonal metric, one would need $b_{23}b_{12}+c_{12}c_{23}k_{||}^2=0,\ g_6 g_5+f_6 f_5=0$. We choose this stronger simplification:
\beq b_{25}=b_{45}=b_{23}=c_{12}=g_5=f_6=0 \ . \label{simpsol2}\eeq
The metric is then:
\beq
g=\left (\begin{array}{cccccc} -\frac{\alpha c_{23} b_{12} k_{\bot}}{k_{||} b_{24}} & 0 & 0 & 0 & 0 & 0 \\ 0 & -\frac{b_{24} b_{12}}{k_{\bot} k_{||} \alpha c_{23}} & 0 & 0 & 0 & 0 \\ 0 & 0 & -\frac{\alpha k_{\bot} c_{23}^3 k_{||}}{b_{12} b_{24}} & 0 & 0 & 0 \\ 0 & 0 & 0 & -\frac{c_{23} k_{||} b_{24} \alpha}{b_{12} k_{\bot}} & 0 & 0 \\ 0 & 0 & 0 & 0 & f_5^2 & 0 \\ 0 & 0 & 0 & 0 & 0 & g_6^2 \end{array}\right )
\eeq
We recover the definite-positiveness of the metric (coherent with (\ref{defpos2})):
\beq \alpha b_{24} c_{23}b_{12}<0 \ . \eeq

Using the same method as before, we then get:
\beq F_3=\frac{e^{-i\theta} (-k_{\bot}^2 c_{23}^2+b_{24}^2)\ |g_6|}{g_s k_{\bot} b_{24}|f_5|} (e^2 \w e^3 \w e^6+\alpha\ e^1 \w e^4 \w e^6)\ , \eeq
\beq d(F_3)=2\frac{e^{-i\theta} (-k_{\bot}^2 c_{23}^2+b_{24}^2)\ |g_6|}{g_s k_{\bot} b_{24}|f_5|}  (e^1 \w e^3 \w e^5 \w e^6 -\alpha\ e^2 \w e^4 \w e^5 \w e^6)\ . \eeq
We see that $F_3$ has two sources, the one along $13$ as expected, and we discover that a second one is then absolutely needed: one along $24$. As before, we compute the co-volumes and get:
\beq V^1=-\frac{k_{\bot}e^{i\theta} f_5g_6 \alpha c_{23}^2}{b_{24}}\ e^1\w e^3 \w e^5 \w e^6,\ V^2=-\frac{e^{i\theta}f_5g_6 b_{24}}{k_{\bot}}\ e^2\w e^4 \w e^5 \w e^6 \ ,\eeq
\beq d(F_3)=-\frac{2(b_{24}^2-k_{\bot}^2 c_{23}^2)\ |g_6| c_{23}^2}{g_s |f_5| C}  (\frac{1}{k_{\bot}^2 c_{23}^2} V^1 -\frac{1}{b_{24}^2} V^2),\ \textrm{with}\ C=f_5g_6 \alpha c_{23}^2 >0 \ .\eeq
The nature of the sources depends on the sign of their charges, which depends here on the value of the parameters. But we can clearly see that there is one O-plane and one D-brane. In both cases, the O-plane is compatible with the manifold. Note also that we clearly have $\sum_i Q_i <0$.\\

The solution with the simple choice of parameters is:
\bea
\textrm{Re}(\Omega_2)_{||} &=& \frac{\alpha k_{\bot}^2 c_{23}^2}{b_{24}} e^1 \w e^3 + b_{24}  e^2 \w e^4 \ ,\nn\\
\textrm{Re}(\Omega_2)_{\bot} &=& b_{12}  e^1 \w e^2-\frac{\alpha c_{23}^2 k_{||}^2}{b_{12}} e^3 \w e^4 \ ,\nn\\
\textrm{Im}(\Omega_2) &=& -\alpha c_{23} e^1 \w e^4+ c_{23} e^2 \w e^3 \ .
\eea
As for the first solution, it is clear from this formulation that what is parallel or orthogonal under $\sigma_{13}$ is also under $\sigma_{24}$. The same goes for $z$ which has only components along $e^5$ and $e^6$. So the solution is clearly compatible with the projections of both sources (in case they are O-planes).

\subsubsection{Third solution}

We proceed as for the previous solutions. We look for IIA solutions on the solvmanifold $s\ 2.5$ which has the following algebra: $(25,-15,\alpha 45,-\alpha 35, 0, 0)$, $\alpha \ \in \ \mathbb{Z}$, but now with an $O6$ in the $136$ directions. We are going to see that this solution is T-dual to the second one, so there will be a lot of similarities between the two. The solution to the usual list of constraints depends on the following real (constant) parameters: $b_{25},b_{45},b_{24},b_{12},b_{23},c_{12},c_{23},f_6,g_5,k_{||},\alpha$ which have to satisfy almost the same conditions as the second solution does: $b_{24}$, $f_6g_5$ and $c_{23}b_{12}-c_{12}b_{23}$ have to be non-zero, $k_{||}$ has to be neither $0$ nor $1$, and $\alpha$ has to be $\pm1$. The general solution is expressed in the usual manner with the following $z^i$:
$$z=i g_5 e^5+f_6 e^6 \ ,$$
$$z^1=(b_{12}+i c_{12}) e^1- (b_{23}+i c_{23}) e^3 -b_{24} e^4 -b_{25} e^5 \ ,$$
$$z^2=e^2 + \frac{\alpha k_{||}^2 (c_{23}b_{12}-c_{12}b_{23})^2}{(b_{12}^2 +c_{12}^2 k_{\bot}^2)(b_{12}+i c_{12})b_{24}}e^3-\frac{\alpha (b_{23}+i c_{23})}{(b_{12}+i c_{12})}e^4 - \frac{1}{b_{24}}\left ( b_{45} + \frac{\alpha b_{25}(b_{23}+i c_{23})}{(b_{12}+i c_{12})} \right ) e^5 \ ,$$
$$b_r=-\frac{k_{||} b_{12} c_{12}}{k_{\bot}(b_{12}^2+c_{12}^2)}\ , \qquad b_i=-\frac{k_{||} b_{12}^2 }{k_{\bot}(b_{12}^2+c_{12}^2)} \ ,$$
\beq t_1=\frac{\alpha k_{||}(c_{23}b_{12}-c_{12}b_{23})}{k_{\bot}b_{24}(b_{12}^2+c_{12}^2)}\ , \qquad
t_2=\frac{b_{24}(b_{12}^2+c_{12}^2 k_{\bot}^2)}{\alpha k_{\bot}k_{||}(c_{23}b_{12}-c_{12}b_{23})}\ .\eeq

The definite-positiveness of the metric is given by
\beq \alpha b_{24} (c_{23}b_{12}-c_{12}b_{23})>0 \ .\label{defpos3}\eeq

For the general solution given before, and given that $\alpha^2=1$, we have $H=0$ and $d(\textrm{Im}(z))=0$. Hence we deduce
\bea
&& F_0 = 0 \ ,\nn\\
&& F_4 = 0 \ .
\eea

The only remaining flux is then $F_2$. The general metric determinant is (clearly non-zero):
\beq \frac{(b_{23}c_{12}-c_{23}b_{12})^4 g_5^2 f_6^2}{(c_{12}^2 k_{\bot}^2+b_{12}^2)^2} \ .\label{detsol3}\eeq

For the same reasons as for the previous solutions, we go to a simpler case. We first choose $b_{25}=b_{45}=0$. Then to get a diagonal metric, one would need $b_{23}b_{12}+c_{12}c_{23}k_{\bot}^2=0$. We choose this stronger simplification:
\beq b_{25}=b_{45}=b_{23}=c_{12}=0 \ .\label{simpsol3}\eeq
The metric is then:
\beq
g=\left (\begin{array}{cccccc} \frac{\alpha c_{23} b_{12} k_{||}}{k_{\bot} b_{24}} & 0 & 0 & 0 & 0 & 0 \\ 0 & \frac{b_{24} b_{12}}{k_{||} k_{\bot} \alpha c_{23}} & 0 & 0 & 0 & 0 \\ 0 & 0 & \frac{\alpha k_{||} c_{23}^3 k_{\bot}}{b_{12} b_{24}} & 0 & 0 & 0 \\ 0 & 0 & 0 & \frac{c_{23} k_{\bot} b_{24} \alpha}{b_{12} k_{||}} & 0 & 0 \\ 0 & 0 & 0 & 0 & g_5^2 & 0 \\ 0 & 0 & 0 & 0 & 0 & f_6^2 \end{array}\right )
\eeq
We recover the definite-positiveness of the metric (coherent with (\ref{defpos3})):
\beq \alpha b_{24} c_{23}b_{12}>0 \ . \eeq

Using the same method as before, we then get:
\beq F_2=\frac{(-k_{||}^2 c_{23}^2+b_{24}^2)\ |f_6|}{g_s k_{||} b_{24}f_6|g_5|} (e^2 \w e^3+\alpha\ e^1 \w e^4) \ ,\eeq
\beq d(F_2)=2\frac{(-k_{||}^2 c_{23}^2+b_{24}^2)\ |f_6|}{g_s k_{||} b_{24}f_6|g_5|}  (e^1 \w e^3 \w e^5 -\alpha\ e^2 \w e^4 \w e^5)\ . \eeq
We see that $F_2$ has two sources, the one along $136$ as expected, and we discover that a second one is then absolutely needed: one along $246$. As before, we compute the co-volumes (see (\ref{Vi})) and get:
\beq V^1=\frac{k_{||} g_5 \alpha c_{23}^2}{b_{24}}\ e^1\w e^3 \w e^5,\ V^2=\frac{g_5 b_{24}}{k_{||}}\ e^2\w e^4 \w e^5 \eeq
\beq d(F_2)=-\frac{2(b_{24}^2-k_{||}^2 c_{23}^2)\ |f_6| c_{23}^2}{g_s |g_5| C} (\frac{1}{k_{||}^2 c_{23}^2} V^1 -\frac{1}{b_{24}^2} V^2),\ \textrm{with}\ C=- g_5f_6 \alpha c_{23}^2 >0 \ .\eeq
The nature of the sources depends on the sign of their charges, which depends here on the value of the parameters. But we can clearly see that there is one O-plane and one D-brane. In both cases, the O-plane is compatible with the manifold. Note also that we clearly have $\sum_i Q_i <0$.\\

The solution with the simple choice of parameters is:
\bea
\textrm{Re}(\Omega_2)_{||} &=& \frac{\alpha k_{||}^2 c_{23}^2}{b_{24}} e^1 \w e^3 + b_{24}  e^2 \w e^4 \ ,\nn\\
\textrm{Re}(\Omega_2)_{\bot} &=& b_{12}  e^1 \w e^2-\frac{\alpha c_{23}^2 k_{\bot}^2}{b_{12}} e^3 \w e^4 \ ,\nn\\
\textrm{Im}(\Omega_2) &=& -\alpha c_{23} e^1 \w e^4+ c_{23} e^2 \w e^3 \ .
\eea
It is clear from this formulation that what is parallel or orthogonal under $\sigma_{136}$ is also under $\sigma_{246}$. The same goes for $\textrm{Re}(z)$ which is along $e^6$ and $\textrm{Im}(z)$ which is along $e^5$. So the solution is clearly compatible with the projections of both sources (in case they are O-planes).\\

We claimed that this solution was T-dual to the second one, with, obviously from the sources, a T-duality in the $e^6$ direction. Note that $e^6$ is a component of $z$. This can be understood the following way. In \cite{Rub}, they derived the T-duality rules for the GCG pure spinors, summed-up in \cite{Scan}. Using these rules, and the $SU(2)$ structure contraction properties (\ref{contrO2}) and (\ref{contrj}), one can show easily that a T-duality in a direction given $\textrm{Re}(z)$ or $\textrm{Im}(z)$ is just the exchange of the two pure spinors (\ref{typeA}) and (\ref{typeB}), modulo a possible phase. This is because the terms in $z$ in the pure spinors get exchanged by the T-duality. The exchange of the pure spinors modulo a phase can be summarized by for instance:
\beq k_{\bot} \rightarrow -k_{||},\ k_{||} \rightarrow k_{\bot} \label{Td} \ . \eeq
Going from one theory to the other, the two pure spinors are always exchanged in the SUSY equations. So with the previous T-duality, a solution in one theory becomes a solution in the T-dual theory. Hence taking a solution in one theory, doing the change (\ref{Td}), one gets a T-dual solution in the other theory, where the T-duality has been done in $\textrm{Re}(z)$ or $\textrm{Im}(z)$ direction. This is exactly what happens between the second and the third solution, that is why we can say they are T-dual. Note that we also understand from (\ref{Td}) why an $SU(3)$ structure is dual to a static $SU(2)$ structure, as it is the case for the solutions in \cite{Scan} (see next subsection).

\subsection{$SU(3)$ or static $SU(2)$ structures limits}\label{Continuation}

In \cite{Scan}, $SU(3)$ or static $SU(2)$ structure solutions were found on the manifolds we have just studied. So it is interesting to see what happens to our solutions when we take one of those two limits: it would be somehow natural to recover the solutions of \cite{Scan}. It was at first the kind of intuition that led us to look for intermediate $SU(2)$ solutions on these manifolds. To take the limit on our solutions, one has two options: taking the limit of the pure spinors, or taking the limit of the structure forms. Taking the limit of the pure spinors might not be a good idea. Indeed, we know pure spinors have different types (see subsection \ref{Puretype}) for each G-structure, so there might be a problem when taking the limit. More precisely, only one of the two spinors keeps the same type in the limit, so this pure spinor might transform smoothly, while the other might not. This is summed-up in this table:
$$\begin{array}{cccccc}  & SU(3) & & \textrm{Int.}\ SU(2) & & \textrm{Stat.}\ SU(2) \\ \Phi_+ & 0 & \longleftarrow & 0 & \dashrightarrow & 2 \\ \Phi_- & 3 & \dashleftarrow & 1 & \longrightarrow & 1 \end{array}$$
with the plain arrows indicating the smooth limits and the dashed ones indicating the limits where there might be a problem. We recover this point when considering the dielectric pure spinors expressions (\ref{su2purespinors}): when one replaces first $j_D$ and $\textrm{Re}(\omega_D)$ by their expressions, and then takes the limit, one does not get the correct expressions for the pure spinors. To get them right, one has to use the following prescription: first take the limit of $j_D$ and $\textrm{Re}(\omega_D)$, and then the limit of the expression obtained.\\

This prescription is more in favor of the second option: taking the limit of the structure forms, and that is what we will do. Looking at the expressions of the dielectric forms $j_D$ and $\omega_D$ in (\ref{dielecform}), we see that their limits give straightforwardly the forms of the limit structures. Actually, we prefer to use the projection basis $\textrm{Re}(\Omega_2)_{||}$, $\textrm{Re}(\Omega_2)_{\bot}$ and $\textrm{Im}(\Omega_2)$, as we gave our solutions with these variables. More precisely, we are going to take the limit of $\textrm{Im}(\Omega_2)$ and $\frac{1}{k_{..}}\textrm{Re}(\Omega_2)_{..}$, where $..$ stands for $||$ or $\bot$. Doing so, we also recover the forms of the limit structures, as one can see from (\ref{link}) or (\ref{rel}). We get\footnote{Note that we recover in these limits the fact that, according to the projection conditions (\ref{proj1}), $j$ and $\textrm{Re}(\Omega_2)$ (and $\textrm{Im}(\Omega_2)$) of the limit structures are the projection eigenbasis.}:
$$\begin{array}{c|c|c}
 & SU(3) & \textrm{Static}\ SU(2)\\
 & k_{\bot}\rightarrow 0 & k_{||}\rightarrow 0 \\
\hline \textrm{IIA} & (\frac{1}{k_{||}}) \textrm{Re}(\Omega_2)_{||}\rightarrow \textrm{Re}(\Omega_2) & (\frac{1}{k_{\bot}}) \textrm{Re}(\Omega_2)_{\bot}\rightarrow \textrm{Re}(\Omega_2) \\
 & \frac{1}{k_{\bot}} \textrm{Re}(\Omega_2)_{\bot}\rightarrow j & -\frac{1}{k_{||}} \textrm{Re}(\Omega_2)_{||}\rightarrow j \\
\hline \textrm{IIB} & (\frac{1}{k_{||}}) \textrm{Re}(\Omega_2)_{\bot}\rightarrow \textrm{Re}(\Omega_2) & (\frac{1}{k_{\bot}}) \textrm{Re}(\Omega_2)_{||}\rightarrow \textrm{Re}(\Omega_2) \\
 & \frac{1}{k_{\bot}} \textrm{Re}(\Omega_2)_{||}\rightarrow j & -\frac{1}{k_{||}} \textrm{Re}(\Omega_2)_{\bot}\rightarrow j
\end{array}
$$
It is clear that $\frac{1}{k_{..}}\textrm{Re}(\Omega_2)_{..}$ is not the best choice for taking the limit since $k_{||}$ or $k_{\bot}$, assumed non-zero, have to go to zero\footnote{The difficulties that can occur are related to the one just explained for the pure spinors, since they both are related to the assumption of $k_{||}$ and $k_{\bot}$ being non-zero.}. Indeed, one can see from the previous arrays that $\Omega_2$ is always recovered smoothly while $j$ is not recovered very easily. For instance in the case IIA and $SU(3)$ limit, $k_{\bot}$ and $\textrm{Re}(\Omega_2)_{\bot}$ both go to zero, and only their fraction is supposed to give back $j$. To get a well-defined limit, we should have a non-zero $j$, and so we must have in the previous example $\textrm{Re}(\Omega_2)_{\bot}\sim k_{\bot} f_2 \rightarrow 0$, where $f_2$ stands for a constant real two-form. Imposing this last condition will give us the behaviour of some of our parameters. It can also sometimes lead to inconsistencies such as the volume form going to zero, and then we can say that there is no limit solution.\\

Here is how we will proceed. By first studying the limit to $j$, we get conditions on the behaviour of our parameters: some go to zero in a specific way, as just explained. Using them, we work out the limit to $\Omega_2$ (extrapolated to $\Omega_3$ in the $SU(3)$ case), and manage to get the $z^i$ of \cite{Scan} solutions, noted $z^i_s$, by factorizing the form as they do. Then we work out completely the limit to $j$ (extrapolated to $J$ in the $SU(3)$ case), and find the needed $t_{is}$ and $b_s$ (same notations as (\ref{factorO2})) to get their solution. Finally, we verify that we have the same fluxes as they do when taking the limit on ours.\\

The validity of this procedure could be discussed further. In particular, we do recover the structure forms found in \cite{Scan} (modulo global normalization factors) as we find maps between their parameters and ours. But there is a possible mismatch for the $H$ flux in the static $SU(2)$ limit, as one can see from its definition in the SUSY conditions (\ref{SUSYA}) or (\ref{SUSYB}). Indeed, if we did not find any $H$ in the intermediate case, we cannot take its limit to recover an $H$ in the static $SU(2)$ limit, while the SUSY conditions allow for a non-trivial $H$ in this limit. This situation will happen for our third solution, as they do find a possible $H$ in \cite{Scan} while we do not. For our second solution, this problem could also have occurred, but no $H$ was found in \cite{Scan}. Note that if there is a mismatch with $H$, then there is possible one with the other fluxes, as we can see from their definitions.

\subsubsection{Limits of the first solution}

Let us first consider the $SU(3)$ limit of the first solution which should correspond to ``Model 1'' of \cite{Scan} (same theory, same manifold, same orientifold(s)). Imposing that $\textrm{Re}(\Omega_2)_{||}$ goes to zero ($\sim k_{\bot}$) and comparing with their $J_s$ gives these behaviours for our parameters: $b_{12}\ll k_{\bot},\ b_{23}\ll k_{\bot}$ and $b_{26} \sim -k_{\bot}$ (with a possible positive constant\footnote{The sign comes from the study of the $C$ appearing in the charges computation and the definite-positiveness of the metric.} that we will not consider for simplicity). Note that a priori in our solution $b_{12}$ and $b_{23}$ could be zero but $b_{26}$ could not, so can we put it to zero? This is actually possible only when taking the limit, we forbade it when looking for solutions because we restricted ourselves to pure intermediate cases. One criteria to verify that the limit is well-defined is that the six-form volume must not go to zero. And $b_{26}$ actually does not appear in it, as one can see from the determinant of the metric (\ref{detsol1}), so it is fine. So using these behaviours of our parameters and the limits given in the array, we get their $\Omega_{3s}$ and $J_s$ with a global normalization difference. The normalization factor affects both $\Omega_3$ and $J$ so that the normalization condition (\ref{SU3comp}) is still satisfied for both our limits and the forms in \cite{Scan}. We decide to take this factor into account by rescaling some of the $z^i_s$ and the $t_{is}$ to match the one we have when taking the limit. We get\footnote{Note that we have here an example of a different choice for the $z^i$, mentioned in subsection \ref{Solconv}. So it would have been surprising to recover theirs by taking the limit of our $z^i$. The way we recovered their solution is a reparametrization, since we computed the two-form in the limit and then refactorized it in the way they did.}:
\beq z^1=(f_1+i f_3) (e^1 -i e^3),\ z^2=e^2 +i \tau e^6,\ z^3=(b_{24}+i c_{24}) (e^4 +i e^5),\ \textrm{with}\ \tau=i \frac{(b_{46}+i c_{46})}{(b_{24}+ic_{24})} \ ,\eeq
\beq t_{1}=1,\ t_{2}=-\frac{1}{\tau_r},\ t_{3}=-\tau_r,\ \textrm{with}\ \tau_r=\textrm{Re}(\tau) \ . \eeq

Looking at our fluxes, we get that $H \rightarrow 0$ as in \cite{Scan}, and we deduce that $F_1 \rightarrow 0$ when we look at the SUSY conditions (\ref{SUSYB}). To compare the $F_3$, we go to the simpler case we chose for our coefficients (\ref{simpsol1}): it gives $\textrm{Im}(\tau)=0$. In this case, we recover $F_1 \rightarrow 0$ when looking at its expression. Moreover, taking the limit on our $F_3$, we recover the solution of \cite{Scan}, once the $t_{is}$ are rescaled as explained.\\

Let us now consider the $SU(2)$ limit. Looking at the condition $\textrm{Re}(\Omega_2)_{\bot} \rightarrow 0$, one gets at least $b_{46}\sim c_{46}\rightarrow 0$ (with a possible constant), and $b_{24}\rightarrow 0$. But this is not allowed, because the volume form would go to zero (see for instance (\ref{detsol1})). So we recover the statement of \cite{Scan}: there is no static $SU(2)$ limit. Note that a T-dual on this manifold to the $SU(3)$ limit would have been a static $SU(2)$ structure with an $O6$. Then, the fact that there is no static $SU(2)$ on this manifold can also be understood by the fact that there is no $O6$ compatible, according to table \ref{Dir}.

\subsubsection{Limits of the second solution}

Let us first consider the $SU(3)$ limit of the second solution. We mention first that no corresponding solution is mentioned in \cite{Scan}. There can be several reasons for this, among them one can be that there is no solution with fluxes which is non T-dual to a warped $T^6$ with an $O3$. We actually do find such a solution, which should be the T-dual to the static $SU(2)$ limit of our third solution (see next subsection). So we will use similar notations. Considering as usual $\textrm{Re}(\Omega_2)_{||}\rightarrow 0\ (\sim k_{\bot})$, we get $b_{25}\sim y_1 k_{\bot},\ b_{45} \sim y_2 k_{\bot},\ b_{24} \sim x k_{\bot}$ with $y_1,\ y_2,\ x$ real constants. As for the previous solution, $b_{25}\rightarrow 0$, $b_{45}\rightarrow 0$ are allowed in our solution, but $b_{24} \rightarrow 0$ is not for an intermediate $SU(2)$ structure. With the same arguments as before, it can actually be allowed in the $SU(3)$ limit (see (\ref{detsol2})). Using these behaviours of our parameters, we get:
\beq \Omega_{3\ SU(3)}=(b_{12}+i c_{12})\ ((f_5+i g_5)\ e^5+(f_6+i g_6)\ e^6) \w (e^1- \tau e^3) \w (e^2 -\alpha \tau e^4 - \frac{1}{x}(y_2 +y_1 \alpha \tau ) e^5) \ ,\eeq
with $\tau=\frac{b_{23}+i c_{23}}{b_{12}+i c_{12}}$ (clearly of the same form as the static $SU(2)$ limit of the third solution).\\

Let us now consider the fluxes. We get $H=0$ and then $F_1=0$. In the simpler case chosen for the parameters (\ref{simpsol2}), we get a non-trivial $F_3$ in the $SU(3)$ limit:
\beq F_{3\ SU(3)} = \frac{e^{-i\theta} (-c_{23}^2+x^2)\ |g_6|}{g_s x |f_5|} (e^2 \w e^3 \w e^6+\alpha\ e^1 \w e^4 \w e^6) \ . \eeq
With this simple choice for the parameters, we have $y_1=y_2=0$, so the solution obtained in the limit is compatible with the two sources appearing when computing the BI.\\

Let us now consider the static $SU(2)$ limit of the second solution, which should correspond to ``Model 2'' in \cite{Scan} (taking $\alpha=1$). Our $z$ is clearly the same as theirs. By imposing that $\textrm{Re}(\Omega_2)_{\bot}$ goes to zero ($\sim k_{||}$) and comparing its limit with their $j_s$, we get these behaviours for our parameters: $b_{12}\sim -x k_{||},\ b_{23} \sim -y k_{||}$ where $x$ and $y$ are real constants. It was forbidden in our solution to put these parameters to zero but when one looks closely at the volume form (see for instance (\ref{detsol2})), one sees it can be allowed in the static $SU(2)$ limit. The solution given in \cite{Scan} is the following:
\beq \Omega_{2s}=(e^1+i(-\tau_2^2 e^2+\tau_2^1 e^4 + \tau_3^1 e^5))\w(e^3 + i(\tau_2^2 e^4 + \tau_3^2 e^5 + (2\frac{b_s}{t_{2s}} \tau_2^2+\frac{t_{1s}}{t_{2s}}\tau_2^1)e^2)) \ ,\eeq
with all parameters real, and $t_{2s}=\frac{1+b_{s}^2}{t_{1s}}$. When taking the limit on our forms, we get the same result, with a global normalization factor difference: our $\Omega_{2\ \textrm{static}\ SU(2)}$ and our $j_{\ \textrm{static}\ SU(2)}$ are obtained by multiplying theirs by $\lambda=\frac{(x c_{23}-y c_{12})^2}{b_{24}(c_{12}^2+x^2)}$. Apart from this normalization, we manage to recover their solution with\footnote{Note that the complicated expressions for the parameters are related to the freedom left in choosing different expressions for the $z^i$ (we did not take the same as them), as mentioned in subsection \ref{Solconv}, and the $t_i$ are different for the same reason.}:
\beq \tau_{2}^2=-\frac{c_{23}}{\lambda},\ \tau_{2}^1=-\frac{2xy c_{23}+c_{12}c_{23}^2-y^2 c_{12}}{(c_{12}^2+x^2) \lambda},\eeq
\beq \tau_{3}^1=\frac{-c_{23} b_{45} (c_{12}^2+x^2) -2 xy b_{25} c_{23}+(y^2-c_{23}^2) b_{25} c_{12} }{b_{24} (c_{12}^2+x^2) \lambda},\ \tau_{3}^2=-\frac{b_{45} c_{12}+b_{25} c_{23}}{b_{24} \lambda} \ ,\eeq
\beq t_{1s}=\frac{c_{12}^2+x^2}{x c_{23} -y c_{12}},\ b_s=-\frac{x y+c_{12} c_{23}}{x c_{23} -y c_{12}} \ . \eeq
Note that this $\lambda$ is a part of the volume obtained in the limit (see (\ref{detsol2})), hence it is well-defined and cannot be zero. Note also that we recover both their $j_s$ and their $\Omega_{2s}$ with a factor $\lambda$ difference, so that the normalization condition (\ref{SU2comp1}) stays correct for us and for them. As this normalization condition implies $\lambda^2$ we have the choice on the sign of the factor in $j$ (we took $+ \lambda$), which is related to the sign of $b_{24}$. It is then related to the sign of the $t_i$ appearing.\\

Let us now look at the fluxes. We have only an $F_3$ as they do. In the simple case chosen for our parameters (\ref{simpsol2}), by taking the limit of our $d(F_3)$, we exactly get theirs, multiplied by $\lambda$ as it should be.

\subsubsection{Limits of the third solution}

We already mentioned that this solution was the T-dual of the second one. In \cite{Scan}, they also mention this point for the limit structures: the $SU(3)$ limit of our solution (with $\alpha=1$) consists of their ``Model 3'', and they mention that it is the T-dual to their ``Model 2'', which is the static $SU(2)$ limit of our second solution, as just discussed. So by this T-duality argument, this $SU(3)$ limit of our solution must match their ``Model 3'', and we will not consider further the $SU(3)$ limit. Note for instance we get the ``same'' (T-dual) limit behaviours of our parameters: $b_{12}\sim x k_{\bot},\ b_{23} \sim y k_{\bot}$ where $x$ and $y$ are real constants.\\

Let us now consider the static $SU(2)$ limit of our third solution, which corresponds to the ``Model 4'' in \cite{Scan}. With the same reasoning, it is probably the T-dual to the $SU(3)$ limit of our second solution, that did not match to any solution found in \cite{Scan}. We first note that our $z$ matches theirs, modulo a global $i$ factor. This difference is due to a different phase convention for the $O6$. Let us look at the other forms. As usual, considering the limit of $\textrm{Re}(\Omega_2)_{||}$ and comparing it to $j_s$ imposes $b_{25}\sim y_1 k_{||},\ b_{45} \sim y_2 k_{||},\ b_{24} \sim x k_{||}$ with $y_1,\ y_2,\ x$ real constants. Once again, $b_{24}$ going to zero can be allowed in this limit (see (\ref{detsol3})).  Using these behaviours of our parameters, we get the solution of \cite{Scan} by taking the limit on our forms. In \cite{Scan} they have:
\beq \Omega_{2s}=(\tau_1^1 e^1+\tau_2^1 e^3)\w(\tau_1^2 e^2 + \frac{\tau_2^1}{\tau_1^1}\tau_1^2 e^4+\tau_3^2 e^5) \ ,\eeq
with complex parameters, and we match it and $j_s$ with:
\beq \tau_1^1=\frac{b_{12}+i c_{12}}{\tau_1^2},\ \tau_2^1=- \frac{b_{23}+i c_{23}}{\tau_1^2},\ \tau_3^2=-\frac{\tau_1^2}{x} \left (y_2 +y_1 \frac{(b_{23}+i c_{23})}{(b_{12}+i c_{12})} \right ) \ ,\eeq
\beq t_{1s}=\frac{(c_{23}b_{12}-b_{23}c_{12}) |\tau_1^2|^2}{(b_{12}^2+c_{12}^2)x}=\frac{1}{t_{2s}} \ ,\eeq
where $\tau_1^2$ is not fixed. Note $t_{1s}t_{2s}=1$ is here the normalization condition (\ref{SU2comp1}).\\

Let us now look at the fluxes. There are slight differences, due to $H$ and $z$. As explained previously, we do not get any $H$, while they do: this is an artefact of our procedure. Note nevertheless that their $H$ is more constrained than it appears to be in \cite{Scan}, once one imposes it to be real. In our simple choice of parameters (\ref{simpsol3}), by taking the limit on our $d(F_2)$, we get exactly theirs, modulo the factor coming from $z$ (related to the difference between our $z$ and theirs), and the following map we have to impose: $|\tau_1^2|^4=c_{23}^2$. This last condition can seem surprising, but this difference is probably related to the absence of $H$ in our limit. Besides, note that in this simplified choice, we get $y_1=y_2=\tau_3^2=0$, hence the solution obtained is clearly compatible with the sources appearing.

\subsection{Some other interesting solutions?}\label{other}

In \cite{Scan} they give the list of all the interesting nil/solvmanifolds and several information about each. Then they checked for each of these manifolds whether there were some $SU(3)$ or static $SU(2)$ solutions. They only found a few, and even less which were ``new'' (not T-dual to a warped $T^6$ with an $O3$). One can ask if we could do the same study for intermediate $SU(2)$ structures. It would be a tedious job, so let us first make a few remarks. We showed that the intermediate $SU(2)$ solutions we already found gave back the solutions found in \cite{Scan} as limit solutions. But these were in \cite{Scan} the only ``new'' solutions. So if there is any other intermediate $SU(2)$ solution on one of the manifolds, there can only be two cases: either this solution has not any well defined limit, or it has but then the limit solution is not ``new''. As an example of the first case, we mention that there might be (to be verified) an intermediate $SU(2)$ solution on $n\ 5.2$ with an orientifold along $56$ which does not seem to have any well defined limit solution, because this set of manifold/orientifold does not appear in the list of solutions of \cite{Scan}. As an example of the second case, let us mention that in \cite{KT} they find an intermediate $SU(2)$ (not ``new'') solution on $n\ 4.4$ and $n\ 4.6$ of which some limit solution was found in \cite{Scan} (and it was T-dual to a warped $T^6$). These remarks point out that to find quickly any other ``new'' intermediate $SU(2)$ solution, we cannot use anymore the same intuition as before: trying to find some on the manifolds where ``new'' $SU(3)$ or static $SU(2)$ solutions were found in \cite{Scan}. We have to use other ideas.\\

To find a ``new'' intermediate $SU(2)$ solution, we will restrict our search to the specific case of solutions with several non completely overlapping orientifolds (in fact there cannot be more than two as we will see). The idea which leads us to do so is that it might be difficult to start with several non-overlapping sources, and get back by T-dualities a single $O3$. Furthermore, we choose the two non-overlapping sources to be orientifolds to use their projection properties: then, one can give some arguments which allow to discard some of the manifolds as candidates for intermediate $SU(2)$ solutions. These arguments also help to understand why a second source was appearing in our solutions, while we were only imposing one. These (technical) arguments are given in appendix \ref{reducset}. Starting with the whole list of possible manifolds and orientifolds, using these arguments (including symmetries of the algebra) we end up with the following restricted set of possible configurations of manifolds/O-planes (with couples of orientifolds between brackets):\\

IIB (and $O5$ sources):
$$n\ 3.3\ (45,16),\ n\ 3.6\ (25,46),\ n\ 3.9\ (25,46),\ n\ 3.13\ (45,26)$$
$$n\ 3.14\ (45,26),\ n\ 4.1\ (26,35),\ n\ 4.2\ (26,35),\ n\ 4.5\ (35,26),\ n\ 4.6\ (35,26)$$
$$s\ 2.2\ (14,23),\ s\ 2.4\ (14,25),\ s\ 2.5\ (13,24),\ s\ 2.6\ (14,23)$$
\beq s\ 3.1\ (14,25) (15,24),\ s\ 3.3\ (13,24),\ s\ 4.1\ (14,25) \ , \eeq

IIA (and $O6$ sources):
$$n\ 3.9\ (235,346),\ n\ 3.10\ (136,235),\ n\ 3.11\ (136,235),\ n\ 3.15\ (235,346),\ n\ 3.16\ (136,235)$$
$$n\ 4.2\ (236,345),\ n\ 4.3\ (146,345),\ n\ 4.4\ (146,345),\ n\ 4.6\ (246,345),\ n\ 4.7\ (135,146) (135,236)$$
$$s\ 2.2\ (135,245),\ s\ 2.5\ (136,246),\ s\ 2.6\ (146,236),\ s\ 3.2\ (146,256)$$
\beq s\ 3.3\ (136,246),\ s\ 3.4\ (145,246),\ s\ 4.1\ (145,246) \ . \eeq

The result is that we do not find any intermediate $SU(2)$ solution with two non completely overlapping orientifolds on any of them\footnote{We notice that there might be a problem with the algebra of $s\ 2.3$ given in \cite{Scan}, because it is supposed to have only two zeros according to its name, and it actually has three zeros. So we did not try anything on it.} for IIB, apart from the previously found solutions: we tried the following configurations (manifold with the tried O-plane in brackets) without success:
$$n\ 3.3 (45),\ n\ 3.6 (46),\ n\ 3.9 (46),\ n\ 3.13 (45),\ n\ 4.1 (26),\ n\ 4.2 (26),\ n\ 4.5 (35),\ n\ 4.6 (35)$$
\beq s\ 2.2 (14),\ s\ 2.4 (14),\ s\ 2.6 (14),\ s\ 3.1 (14)(15),\ s\ 3.3 (13),\ s\ 4.1 (14) \ . \eeq
For IIA, the work still has to be done.

\section{Conclusion}\label{Concl}

In this paper, we have looked for ``new'' supersymmetric four-dimensional Minkowski flux vacua of type II string  theory, with intermediate $SU(2)$ structure. They are ``new'' in the sense they are not T-dual to a $T^6$ with an $O3$. We found three of them, in the large volume limit with smeared sources and constant parameters. Two of them are T-duals among themselves. To find these vacua, we introduced a new $SU(2)$ structure, that transforms simply under the orientifold projection, and which actually corresponds to the $SU(2)$ structure appearing with the dielectric pure spinors. Using these variables, we rewrote the projection conditions given in \cite{KT} in a more tractable way, and at the same time, the SUSY conditions became much simpler to solve. On the solutions found, we took the limit to the $SU(3)$ or the static $SU(2)$ cases, and recovered the solutions of \cite{Scan}, hence getting some intuition on what a dynamical $SU(3) \times SU(3)$ structure could look like.\\

Some points remain to be studied. One interesting point is the number $\mathcal{N}$ of four-dimensional SUSY preserved by the vacuum. Since these manifolds are parallelizable, the effective action  is a priori  maximally supersymmetric. Part of the supersymmetry can be broken by the presence of sources. Then, the vacuum can only preserve a fraction $\mathcal{N}$ of it. In \cite{Scan} and \cite{KT}, $\mathcal{N}$ was given in simple cases. Generically, looking at (\ref{10dspinoransatz}), one has to count the number of different pairs of internal spinors which are solutions to the SUSY conditions, and give the same vacuum. In other words, one has to count the number of different pairs of pure spinors which are solutions, and give the same metric and fluxes. It is the same as identifying the freedom left in a generic solution, which is for general solutions not an easy thing to do. That is why we did not discuss it in this paper, but a careful study could be interesting.\\

Another point is applying these techniques to study the possibility of AdS vacua with intermediate $SU(2)$ structures. Actually, after the first appearance of this paper, it was shown in \cite{CKKLTZ} that such solutions cannot exist.\\

A last point to study is the appearance of a second source in our solutions, while we were only imposing one. If we knew in advance that a second source was going to be present, this could have simplified the search in the case of an O-plane because of the other projection conditions to impose. In subsection \ref{other}, we discussed why the second O-plane could appear at the same time, but it is not clear whether its presence is necessary.\\

Finding these ``new'' solutions has several interests. It provides new examples of vacua on GCY, not related to the usual and widely studied $T^6$. It then gives some insight on new corners of the landscape, providing for instance new set-ups to compactify and find low energy effective actions. The compactification on these new manifolds has already been studied, and some arguments to find the effective actions have been given \cite{Reduc, MPT, Mart, Dav}. Note that finding first the four-dimensional effective action and then its vacuum has been proved to be equivalent to find directly the ten-dimensional vacuum on the product space-time the way we did here \cite{Mart, Dav}.\\

Another possible interest is the link with non-geometrical backgrounds, as done in \cite{Scan}. In particular, it was mentioned in \cite{Scan} a possible link due to an asymmetric orbifold \cite{HMW, DH}. These new solutions might provide new ingredients to understand it.\\

Finally, the formalism developed here could be interesting for dynamical solutions. Indeed, a dynamical solution would generically have the form of an intermediate $SU(2)$ structure solution everywhere on the internal manifold, except at some points where it becomes an $SU(3)$ structure (or a static $SU(2)$ structure). In this paper we showed that the dielectric pure spinors and the associated $SU(2)$ structure were the good variables in which to find intermediate $SU(2)$ structure solutions, so they are probably the best variables to find dynamical solutions. But so far, despite the simplicity of the equations, our efforts in this direction have not met success.\\

An open side question concerns the search of a better discrimination of the manifolds on which to find a supersymmetric flux vacuum. Indeed, in \cite{Scan}, they looked among a long list of GCY and only found a few on which there was some vacuum. In the same way, in this paper, we tried to find vacua on some other manifolds without success. This seems to indicate the existence of some other refined criteria for which manifold to use, that we are missing at the moment. The mathematical specification of the manifolds is known: as explained, the manifolds should be a twisted GCY admitting an $SU(3)\times SU(3)$ structure, and being compatible with at least one O-plane. The existence of the $SU(3)\times SU(3)$ structure, and whether this structure is compatible with the orientifold projection, might be for instance criteria that haven't been implemented before beginning the search for vacua in \cite{Scan} and in this paper. These could lead to a restricted set of manifolds/orientifolds.\\

\textbf{Acknowledgments}\\

I would like to thank my PhD supervisor Michela Petrini for her help in many aspects all along this work. I would also like to thank D. Cassani, M. Gra\~na, N. Halmagyi, P. Koerber, R. Minasian, and A. Tomasiello, for helpful exchanges and interesting discussions. I acknowledge partial support by the RTN contract MRTN-CT-2004-512194 and by ANR grant BLAN05-0079-01.

\appendix

\section{Conventions and derivation of background formulas}\label{conv}

\subsection{Some conventions of differential forms}\label{Convdiff}

In this appendix we give our conventions on (internal) gamma matrices, differential forms, some useful formulas about contractions, and conventions for the (six-dimensional) Hodge star.\\

We choose hermitian $\gamma$ matrices (they are all purely imaginary and antisymmetric): $\gamma^{i\dag}=\gamma^i$.\\

Here are some identities used (see \cite{Gamma} for more):
\bea
\{\gamma^m,\gamma^n\}=2g^{mn} && [\gamma^m,\gamma^n]=2\gamma^{mn} \nn\\
\{\gamma^{mn},\gamma^p\}=2\gamma^{mnp} && [\gamma^{mn},\gamma^p]=-4\delta^{p[m}\gamma^{n]} \nn\\
\{\gamma^{mnpq},\gamma^r\}=2\gamma^{mnpqr} && [\gamma^{mnpq},\gamma^r]=-8\delta^{r[m}\gamma^{npq]} \ .
\eea

We take as a convention for a $p$-form $A$:
\beq \gamma^{\mu_1...\mu_p}\leftrightarrow dx^{\mu_1}\w ... \w dx^{\mu_p} \qquad A=\frac{1}{p!}A_{\mu_1...\mu_p}\gamma^{\mu_1...\mu_p} \ . \eeq

With some abuse in the notation, when we write the conjugate of a form expressed with real indices (i.e. on a real basis), we mean the conjugate of its components, hence for the one-form $z$ appearing in the main part, we have ($\mu$ being a real index):
\beq \overline{z}=\overline{z}_\mu  \gamma^\mu \ . \eeq

For a $p$-tensor $A$, we define the antisymmetrization (with the $p!$ possible terms on the right-hand side) as:
\beq A_{[\mu_1...\mu_p]}=\frac{1}{p!}(A_{\mu_1\mu_2\mu_3...\mu_p}-A_{\mu_2\mu_1\mu_3...\mu_p}+A_{\mu_2\mu_3\mu_1...\mu_p}+... + A_{\mu_3\mu_4\mu_1\mu_2\mu_5...\mu_p}+...)\ . \eeq
For a $p$-form $A$ and $q$-form $B$, we have the convention:
\beq \frac{1}{(p+q)!}(A\w B)_{\mu_1...\mu_{p+q}}=\frac{1}{p!q!}A_{[\mu_1...\mu_p} B_{\mu_{p+1}...\mu_{p+q}]} \ .\eeq

For a $p$-form $A$ and a $1$-form $b=b_i\gamma^i$, we define the contraction:
\beq b\llcorner A=\frac{1}{p!}b^{\nu} A_{\mu_1...\mu_p}\ p\  \delta_{\nu}^{[\mu_1}\gamma^{\mu_2...\mu_p]}=\frac{1}{(p-1)!}b^{\mu_1} A_{\mu_1...\mu_p} \gamma^{\mu_2...\mu_p} \ . \label{Contraction}\eeq
For generic $1$-form $x$, $p$-form $A$ and $q$-form $B$, one has:
\beq x\llcorner (A\w B) = (x\llcorner A) \w B + (-1)^p\ A \w (x\llcorner B) \label{contrac2}\ .\eeq

We now give the conventions for the Hodge star $*$, with a given metric $g$. We introduce the totally antisymmetric tensor $\epsilon$ by $\epsilon_{\mu_1..\mu_m}=+1/-1$ for $(\mu_1..\mu_m)$ being any even/odd permutation of $(1..m)$, and $0$ otherwise. Then, the convention used for the Hodge star is\footnote{We take the same ``awkward sign convention'' as in \cite{Scan}, in order to use the same pure spinors SUSY equations and the same calibration of the sources.}:
\beq *(dx^{\mu_1} \w ... \w dx^{\mu_k})= \frac{\sqrt{|g|}}{(n-k) !} (-1)^{(n-k)k}\ \epsilon^{\mu_1 .. \mu_k\ \mu_{k+1} .. \mu_n}\ g_{\mu_{k+1} \nu_{k+1}} .. g_{\mu_n \nu_n}\ dx^{\nu_{k+1}} \w ... \w dx^{\nu_n} \ , \eeq
with $n$ the dimension of the space, $|g|$ the determinant of the metric. In the eigenvector basis $(v^1, .., v^n)$, with diagonalized metric $D$, we get for a $k$-form:
\beq *(v^{\mu_1} \w .. \w v^{\mu_k})= (-1)^{(n-k)k}\ \frac{\epsilon_{\mu_1 .. \mu_n}}{\sqrt{|g|}} D_{\mu_{k+1} \mu_{k+1}} .. D_{\mu_{n} \mu_{n}} v^{\mu_{k+1}} \w .. \w v^{\mu_n} \ , \eeq
without any summation on $\mu_{k+1}$, .., $\mu_n$, as we took off the $(n-k)!$, i.e. these indices are fixed; the $\epsilon_{\mu_1 .. \mu_n}$ is then only there for a sign. Note for a $p$-form $A_p$, one has:
\beq **A_p= (-1)^{(n-p)p}\ A_p= (-1)^{(n-1)p}\ A_p \ . \eeq

\subsection{$SU(2)$ structure conditions}\label{SU(2)}

In this appendix we derive in a specific way the $SU(2)$ structure conditions given in subsection \ref{strucgpprop}. We start by considering a globally defined spinor $\eta_+$: this gives an $SU(3)$ structure which has the properties (\ref{SU3comp}). Let us now assume there is some holomorphic globally defined one-form $z$, for which we recall $||z||^2=\overline{z}\llcorner z=z \llcorner \overline{z}=\overline{z}^{\mu}z_{\mu}=2$. One can then always define two-forms from it:
\beq j=J-\frac{i}{2}z\w \overline{z} \qquad \Omega_2=\frac{1}{2}\overline{z}\llcorner \Omega_3 \ . \label{SU2ap}\eeq
Note that $j$ is clearly real. We are going to show that these define an $SU(2)$ structure (the one naturally embedded in the $SU(3)$) since they satisfy the conditions (\ref{SU2comp1}), (\ref{SU2comp2}), and (\ref{SU2comp3}).\\

Holomorphicity is defined with respect to the almost complex structure (see footnote \ref{holo}). Then, one can always have an hermitian metric (its non-zero components have one index holomorphic and the other anthropomorphic). Using this metric and some holomorphicity arguments in six dimensions, we first get that $z\llcorner \Omega_3=0$, $z\llcorner z=\overline{z} \llcorner \overline{z}=0$. Furthermore, we get that $\Omega_2$ is holomorphic, and deduce the following structure conditions:
\beq \Omega_2\w \Omega_2 =0 \ , \label{OO} \eeq
\beq z\llcorner \Omega_2 =0,\ \overline{z} \llcorner \Omega_2 =0 \label{contrO2}\ .\eeq
Using the same arguments, we get that $z\w\Omega_3 =0$, and using (\ref{contrac2}), we have: $0=\overline{z}\llcorner (z\w\Omega_3) = 2 \Omega_3 - z\w (\overline{z}\llcorner \Omega_3)$, hence
\beq \Omega_3= z\w \Omega_2 \label{def2O2ap}\ .\eeq

Let us now recover the structure conditions involving $j$. We get using (\ref{contrac2}): $z\llcorner (\frac{z\w\overline{z}}{2})=-z$, $\overline{z}\llcorner (\frac{z\w\overline{z}}{2})=\overline{z}$. We have (using our almost complex structure and real indices) $\overline{z}\llcorner J = i\overline{z}$, because
\beq(\overline{z}\llcorner J)_{\nu}=\overline{z}^{\mu} J_{\mu\nu}=-J_{\nu\mu}\overline{z}^{\mu}=-J_{\nu}^{\ \ \mu}\ \overline{z}_{\mu}=-(-i)\overline{z}_{\nu}=i\overline{z}_{\nu}\ .\eeq
So we deduce from the definition of $j$ the following structure conditions:
\beq\overline{z}\llcorner j =0,\ z\llcorner j =0 \ . \label{contrj}\eeq
Using $J\w \Omega_3=0$ and (\ref{def2O2ap}), we deduce $z\w j\w \Omega_2=0$, and using (\ref{contrac2}), we then get:
\beq j\w \Omega_2=0 \ .\label{jO}\eeq

To recover the remaining structure condition (\ref{SU2comp1}), we express the equality $\frac{4}{3}J^3= i \Omega_3\w \overline{\Omega}_3 $ in terms of $z$, $j$ and $\Omega_2$, and get $\frac{4}{3}(j+\frac{i}{2}z\w \overline{z})^3= i z\w \overline{z}\w \Omega_2\w \overline{\Omega}_2  $.
Then, using the previously derived properties, contracting last formula with $z$ and then contracting with $\overline{z}$, we finally get:
\beq 2\ j^2= \Omega_2\w \overline{\Omega}_2  \ .\label{SU2vol}\eeq

Going back to $\frac{4}{3}J^3= i \Omega_3\w \overline{\Omega}_3 $, one deduces with (\ref{SU2vol}):
\beq j^3=0\ .\label{j3}\eeq

\subsection{Details on the compatibility conditions}\label{proofcom}

In subsection \ref{Puretype}, we explained that we needed a pair of compatible pure spinors. We mentioned that the compatibility conditions were actually implied by a set of $SU(2)$ structure conditions seen in subsection \ref{strucgpprop}. We are going to prove this implication here. The $SU(2)$ structure conditions involved are (\ref{OO}), (\ref{jO}), (\ref{SU2vol}), and (\ref{j3}). We will use the formulas (\ref{purespin}) for the pure spinors, which are valid for any structure (intermediate or static $SU(2)$, $SU(3)$), hence this result is valid for any structure. We give the following useful formula for any $p$-form $A_p$ and $q$-form $B_q$:
\beq \lambda(A_p\w B_q)=(-1)^{pq}\lambda(A_p)\w \lambda(B_q) \ ,\label{lambda2}\eeq
and we recall the compatibility conditions given in subsection \ref{Puretype} (with the $\Phi_i$ defined in (\ref{Phi12pm})):
\beq \left\langle \Phi_1, \overline{\Phi}_1  \right\rangle=\left\langle \Phi_2, \overline{\Phi}_2  \right\rangle \neq 0 \ ,\eeq
\beq \left\langle \Phi_1, X \cdot \Phi_2 \right\rangle=\left\langle \overline{\Phi}_1 , X \cdot \Phi_2 \right\rangle=0,\ \forall\ X=(x,y)\ \in\ T \oplus T^* \ .\eeq
In the following, we will use the $\Phi_i$ defined in (\ref{Phi12pm}) for IIA, but note these conditions are actually independent of the theory, since they are only involving a generic pair of pure spinors.\\

Using (\ref{purespin}) for the pure spinors, the first compatibility condition gives
\beq z\w \overline{z}\w \left( 2k_{\bot}^2 j^2+k_{||}^2 \Omega_2\w \overline{\Omega}_2 -2k_{||}k_{\bot}j\w \textrm{Re}(\Omega_2) \right) \neq 0 \ ,\label{com0}\eeq
\beq k_{||}^2 \frac{4}{3}ij^3 +ik_{||}k_{\bot} j^2\w \textrm{Re}(\Omega_2)=4\frac{z\w\overline{z}}{||z||^2}\w \left( j^2(k_{||}^2-k_{\bot}^2)+ \frac{1}{2} \Omega_2\w \overline{\Omega}_2  (k_{\bot}^2-k_{||}^2) + 2 j\w \textrm{Re}(\Omega_2) k_{||}k_{\bot}\right) \label{com1}\ .\eeq
One can see that imposing (\ref{jO}), (\ref{SU2vol}) and (\ref{j3}), (\ref{com1}) is automatically satisfied. Only (\ref{com0}) remains to be satisfied; it corresponds to the volume form being non-zero.\\

Let us now focus on the second compatibility condition. Since this condition is valid for any $X$, it is sufficient to study it in the two different cases where $X=(x,0)$ and $X=(0,y)$. Then let us first look at $X=(0,y)$ and the condition $\left\langle \Phi_1, X \cdot \Phi_2 \right\rangle=0$. One gets:
\beq y\w z\w\Omega_2\w(k_{||}k_{\bot}\Omega_2+(k_{||}^2-k_{\bot}^2)j)=0 \label{com2}\ .\eeq
As (\ref{com2}) is valid for any $y$, we get:
\beq z\w \Omega_2\w(k_{||}k_{\bot}\Omega_2+(k_{||}^2-k_{\bot}^2)j)=0\ .\eeq
If one imposes (\ref{OO}) and (\ref{jO}), (\ref{com2}) is automatically satisfied.\\

Let us now consider $X=(x,0)$ and still $\left\langle \Phi_1, X \cdot \Phi_2 \right\rangle=0$. Using (\ref{contrac2}) and the following useful formula valid $\forall x\ \in\ T,\ \forall n\ \epsilon \ \mathbb{N}^*$
\beq x\llcorner j^n = n\ j^{(n-1)} \w (x\llcorner j) \ , \label{jn}\eeq
one gets the following top form in terms of $x\llcorner z$, $x\llcorner j$, and $x\llcorner \Omega_2$:
\beq (x\llcorner z) \left ( \frac{i}{2} \Omega_2 \w j^2 + \frac{z\w \overline{z}}{2} \Omega_2\w (-k_{||}k_{\bot} \Omega_2 + j (k_{\bot}^2-k_{||}^2)) \right ) - z\w \left (i k_{\bot}^2 \Omega_2 \w j\w (x\llcorner j) + \frac{k_{||}^2}{2} j^2 \w (x\llcorner \Omega_2) \right )\ .\eeq
Apart from the term in $j^2 \w (x\llcorner \Omega_2)$, the previous expression is obviously zero when one imposes (\ref{OO}) and (\ref{jO}). Using (\ref{contrac2}) and (\ref{jn}), one has
\beq x\llcorner (j^2 \w\Omega_2)=2\ j\w x\llcorner (j) \w\Omega_2+j^2 \w (x\llcorner \Omega_2)\ .\eeq
Hence the term in $j^2 \w (x\llcorner \Omega_2)$ is also zero when using (\ref{jO}), so the whole expression vanishes with (\ref{OO}) and (\ref{jO}). Thus, $\left\langle \Phi_1, X \cdot \Phi_2 \right\rangle=0$ is automatically satisfied for any $X$ when (\ref{OO}) and (\ref{jO}) are imposed.\\

One can play the same game with the condition $\left\langle \overline{\Phi}_1 , X \cdot \Phi_2 \right\rangle=0$. For $X=(0,y)$, one gets:
\beq y\w z\w \left ( k_{||}k_{\bot} (\Omega_2\w \overline{\Omega}_2  - 2\ j^2) +j\w (k_{||}^2 \Omega_2-k_{\bot}^2\overline{\Omega}_2  ) \right ) =0 \ ,\eeq
which is obviously satisfied by imposing (\ref{jO}) and (\ref{SU2vol}). For the $X=(x,0)$ case, one gets:
\bea
(x\llcorner z) \left (-\frac{4i}{3}k_{||}k_{\bot}j^3 +\frac{i}{2} j^2\w (k_{||}^2\Omega_2 -k_{\bot}^2 \overline{\Omega}_2 ) - \frac{z\w \overline{z}}{2}\w (k_{||}k_{\bot} (\Omega_2\w \overline{\Omega}_2  -2 j^2 ) + j \w (k_{||}^2 \Omega_2 -k_{\bot}^2\overline{\Omega}_2 )) \right ) && \nn \\
+ z\w \left (i k_{\bot} (x\llcorner j) \w j\w  (2 k_{||} j + k_{\bot} \overline{\Omega}_2 ) - \frac{k_{||}^2}{2} j^2 \w (x\llcorner \Omega_2) \right )=0 \ . \quad &&
\eea
Using the same kinds of tricks as before ((\ref{j3}) gives $j^2\w (x\llcorner j)=0$), we get that (\ref{jO}), (\ref{SU2vol}) and (\ref{j3}) imply that the whole expression is zero. Thus, $\left\langle \overline{\Phi}_1 , X \cdot \Phi_2 \right\rangle=0$ is automatically satisfied for any $X$ when (\ref{jO}), (\ref{SU2vol}) and (\ref{j3}) are imposed.\\

We add the following point referring to subsection \ref{projbasis}: (\ref{com1}) can be rewritten in terms of the projection basis variables. It gives an equation which can be decomposed in the two following equations after projection:
\bea
\textrm{Re}(\Omega_2)_{||}^3 \frac{\gamma^3}{6}(3+r(3+2k_{||}^2))-\textrm{Re}(\Omega_2)_{\bot}^2\w \textrm{Re}(\Omega_2)_{||} \frac{1}{2\gamma}(1-r(3+2k_{||}^2))\ \  && \nn\\
+\frac{8}{||z||^2}\frac{1+r}{2} \textrm{Re}(z)\w \textrm{Im}(z)\w (\frac{\textrm{Re}(\Omega_2)_{||}^2}{1-\cos(2\phi)}-\frac{\textrm{Re}(\Omega_2)_{\bot}^2}{1+\cos(2\phi)})=0 \ , && \label{comb01}\\
\nn\\
\textrm{Re}(\Omega_2)_{\bot}^3 \frac{1}{6\gamma^3}(3-r(3+2k_{||}^2))-\textrm{Re}(\Omega_2)_{||}^2\w \textrm{Re}(\Omega_2)_{\bot} \frac{\gamma}{2}(1+r(3+2k_{||}^2))\ \  && \nn\\
+\frac{8}{||z||^2}\frac{1-r}{2}\textrm{Re}(z)\w \textrm{Im}(z)\w (\frac{\textrm{Re}(\Omega_2)_{||}^2}{1-\cos(2\phi)}-\frac{\textrm{Re}(\Omega_2)_{\bot}^2}{1+\cos(2\phi)})=0 \ . && \label{comb02}
\eea
Actually, one can show that these two equations are automatically satisfied using (\ref{com5}) and (\ref{com6}) which are a rewriting of some $SU(2)$ structure conditions, because each of the three terms is zero. So we recover the fact that (\ref{com1}) is automatically satisfied after imposing the $SU(2)$ structure conditions.

\section{Going to the projection basis}\label{Projbasap}

In subsection \ref{purespindiel}, we explained that the good variables to use were the projection basis:
\beq \textrm{Re}(z),\ \textrm{Im}(z),\ \textrm{Im}(\Omega_2),\ \textrm{Re}(\Omega_2)_{||},\ \textrm{Re}(\Omega_2)_{\bot},\ (j_{||},\ j_{\bot}) \ ,\eeq
where $j_{||},\ j_{\bot}$ can eliminated using the projection conditions (\ref{projb1}). So in this appendix, we rewrite the different equations to be solved in terms of these variables.

\subsection{$SU(2)$ structure conditions}\label{projbasis}

In this appendix we rewrite the $SU(2)$ structure conditions implying the compatibility conditions (see appendix \ref{proofcom}), namely (\ref{OO}), (\ref{jO}), (\ref{SU2vol}) and (\ref{j3}). To do so we also use the projection conditions (\ref{projb1}). The $SU(2)$ structure conditions (\ref{OO}) and (\ref{jO}) are equivalent to (for both theories):
\beq \textrm{Im}(\Omega_2) \w \textrm{Re}(\Omega_2)_{||}=0\label{com3} \ ,\eeq
\beq \textrm{Im}(\Omega_2) \w \textrm{Re}(\Omega_2)_{\bot}=0\label{com4} \ ,\eeq
\beq \textrm{Re}(\Omega_2)_{||} \w \textrm{Re}(\Omega_2)_{\bot}=0\label{com5} \ ,\eeq
\beq \textrm{Re}(\Omega_2)_{||}^2 = \frac{1}{\gamma^2} \textrm{Re}(\Omega_2)_{\bot}^2\label{com6} \ ,\eeq
\beq \textrm{Re}(\Omega_2)_{||}^2+ \textrm{Re}(\Omega_2)_{\bot}^2 = \textrm{Im}(\Omega_2)^2 \label{com7}\ .\eeq
We do not get any new condition from (\ref{SU2vol}) and (\ref{j3}), which can be understood the following way: as discussed in subsection \ref{purespindiel}, $z,\ \textrm{Im}(\Omega_2),\ \textrm{Re}(\Omega_2)_{||},\ \textrm{Re}(\Omega_2)_{\bot}$ defines, modulo a rescaling, a new $SU(2)$ structure (obtained by a rotation from the previous one). And so it is natural \cite{DallAg} to have the five previous ``wedge conditions'', and only them.\\

We recall that this last set of conditions, together with the projection conditions, is then enough to get all the compatibility conditions except from (\ref{com0}). For instance, in appendix \ref{proofcom}, we rewrote the compatibility condition (\ref{com1}) in terms of the projection basis variables and show that it was automatically satisfied using (\ref{com5}) and (\ref{com6}). Using the last relations and the projection basis, we can also rewrite (\ref{com0}):
$$ \textrm{Re}(z)\w \textrm{Im}(z)\w(\textrm{Im}(\Omega_2)^2+\frac{1}{k_{||}^2}(\frac{1-r}{2}\textrm{Re}(\Omega_2)_{||}^2+\frac{1+r}{2}\textrm{Re}(\Omega_2)_{\bot}^2))\neq0 \ ,$$
\beq \Leftrightarrow  \textrm{Re}(z)\w \textrm{Im}(z)\w \textrm{Re}(\Omega_2)_{||}^2 \neq0 \ .\label{comb1}\eeq

\subsection{SUSY conditions}\label{SUSYapprincipal}

We derive in this appendix the SUSY conditions, starting from (\ref{susy1}), (\ref{susy2}), and (\ref{susy3}) and a general expressions for the pure spinors, and then explaining the various steps leading to the equations given in subsection \ref{SUSYsec}.

\subsubsection{SUSY conditions derivation}\label{SUSYap}

We first use the following general expressions for the pure spinors:
\bea
\Phi_+ &=& \frac{a\overline{b}}{8}N^2 e^{\frac{1}{||z||^2} z\w \overline{z}} (k_{||}e^{-ij}-ik_{\bot}\Omega_2) \ ,\nn \\
\Phi_- &=& -\frac{ab}{8}\frac{\sqrt{2}}{||z||}N^2 z\w (k_{\bot}e^{-ij}+i k_{||}\Omega_2) \ ,
\eea
with $a,\ b,\ ||z||,\ N=||\eta_+||$ constant and non-zero, and $k_{||},\ k_{\bot}$ constant, and without any further fixing. For IIA, we just choose $ab$ real (as it is the case when fixing further) and for IIB, we choose $a\overline{b}$ real, as it is the case for the $O5$ projection. We recall that the fluxes are real. We then get the following equations, where (\ref{susy1}) has been decomposed under its real and imaginary parts:
\bea
\textrm{IIA}&:& F_6=0 \nn\\
& & d(e^{2A-\phi})k_{||}=0 \nn\\
& & d(e^{A-\phi}\textrm{Re}(z))k_{\bot}=0 \nn\\
& & d(e^{3A-\phi}\textrm{Im}(z))k_{\bot}=-\frac{e^{4A}||z||}{\sqrt{2}abN^2}*F_4 \nn\\
& & d(e^{2A-\phi}\textrm{Im}(\Omega_2))k_{\bot}=e^{2A-\phi}k_{||}H \nn\\
& & d(e^{2A-\phi}(-k_{||}j-k_{\bot}\textrm{Re}(\Omega_2)+k_{||}\frac{z\w \overline{z}}{i||z||^2}))=0 \nn\\
& & d(e^{A-\phi}(-k_{||}\textrm{Re}(z)\textrm{Im}(\Omega_2)-\textrm{Im}(z)(-k_{\bot}j+k_{||}\textrm{Re}(\Omega_2))))=e^{A-\phi}k_{\bot}H\textrm{Re}(z) \nn\\
& & d(e^{3A-\phi}(-k_{||}\textrm{Im}(z)\textrm{Im}(\Omega_2)+\textrm{Re}(z)(-k_{\bot}j+k_{||}\textrm{Re}(\Omega_2))))-e^{3A-\phi}k_{\bot}H\textrm{Im}(z)=\frac{e^{4A}||z||}{\sqrt{2}abN^2}*F_2 \nn\\
& & d(e^{2A-\phi}(-\frac{1}{2}k_{||}j^2-\frac{z\w \overline{z}}{i||z||^2}(-k_{||}j-k_{\bot}\textrm{Re}(\Omega_2))))=e^{2A-\phi}k_{\bot} H\w \textrm{Im}(\Omega_2) \nn\\
& & d(e^{2A-\phi}\frac{z\w \overline{z}}{i||z||^2}\w \textrm{Im}(\Omega_2))k_{\bot}=e^{2A-\phi}H(-k_{||}j-k_{\bot}\textrm{Re}(\Omega_2)+k_{||}\frac{z\w \overline{z}}{i||z||^2}) \nn\\
& & \frac{1}{2}k_{\bot} \textrm{Re}(z)\w d(j^2)=H(-k_{||}\textrm{Re}(z)\textrm{Im}(\Omega_2)-\textrm{Im}(z)(-k_{\bot}j+k_{||}\textrm{Re}(\Omega_2))) \nn\\
& & d(e^{3A-\phi}(-\frac{1}{2}j^2\w \textrm{Im}(z)))k_{\bot}-e^{3A-\phi}H\w(-k_{||}\textrm{Im}(z)\textrm{Im}(\Omega_2)+\textrm{Re}(z)(-k_{\bot}j+k_{||}\textrm{Re}(\Omega_2))) \nn\\
& & \qquad \qquad \qquad \qquad \qquad \qquad \qquad \qquad \qquad \qquad \qquad \qquad \qquad \qquad =-\frac{e^{4A}||z||}{\sqrt{2}abN^2}*F_0\ , \label{SUSYapA}
\eea
\bea
\textrm{IIB}&:& F_5=0 \nn\\
& & d(e^{A-\phi})k_{||}=0 \nn\\
& & d(e^{2A-\phi}\textrm{Re}(z))k_{\bot}=0 \nn\\
& & d(e^{2A-\phi}\textrm{Im}(z))k_{\bot}=0 \nn\\
& & d(e^{A-\phi}\textrm{Im}(\Omega_2))k_{\bot}=e^{A-\phi}k_{||}H \nn\\
& & d(e^{3A-\phi}(-k_{||}j-k_{\bot}\textrm{Re}(\Omega_2)+k_{||}\frac{z\w \overline{z}}{i||z||^2}))=-\frac{e^{4A}}{a\overline{b}N^2}*F_3 \nn\\
& & d(e^{2A-\phi}(-k_{||}\textrm{Re}(z)\w \textrm{Im}(\Omega_2)-\textrm{Im}(z)\w (-k_{\bot}j+k_{||}\textrm{Re}(\Omega_2))))=e^{2A-\phi}k_{\bot}H\w \textrm{Re}(z) \nn\\
& & d(e^{2A-\phi}(\textrm{Re}(z)\w (-k_{\bot}j+k_{||}\textrm{Re}(\Omega_2))-k_{||}\textrm{Im}(z)\w \textrm{Im}(\Omega_2)))=e^{2A-\phi}k_{\bot}H\w \textrm{Im}(z) \nn\\
& & d(e^{A-\phi}(-\frac{1}{2}k_{||}j^2+\frac{z\w \overline{z}}{i||z||^2}(k_{||}j+k_{\bot}\textrm{Re}(\Omega_2))))=e^{A-\phi}k_{\bot}H\w \textrm{Im}(\Omega_2) \nn\\
& & d(e^{3A-\phi}k_{\bot}\frac{z\w \overline{z}}{i||z||^2}\w \textrm{Im}(\Omega_2))-e^{3A-\phi}H\w(-k_{||}j-k_{\bot}\textrm{Re}(\Omega_2)+k_{||}\frac{z\w \overline{z}}{i||z||^2})=\frac{e^{4A}}{a\overline{b}N^2}*F_1 \nn\\
& & \frac{1}{2}k_{\bot}\textrm{Re}(z)\w d(j^2)=H\w (-k_{||}\textrm{Re}(z)\w \textrm{Im}(\Omega_2)-\textrm{Im}(z)\w (-k_{\bot}j+k_{||}\textrm{Re}(\Omega_2))) \nn\\
& & \frac{1}{2}k_{\bot}\textrm{Im}(z)\w d(j^2)=H\w (\textrm{Re}(z)\w (-k_{\bot}j+k_{||}\textrm{Re}(\Omega_2))-k_{||}\textrm{Im}(z)\w \textrm{Im}(\Omega_2))\ . \label{SUSYapB}
\eea

Then, one goes further by fixing as usual the parameters ($a=\overline{b}$ and $b=ae^{i\theta}$, $N=||\eta_+||=1$ and $||z||^2=2$), going to the large volume limit (see subsection \ref{SUSYsec}), and assuming $k_{||},\ k_{\bot}$ to be non-zero. The next step is to introduce the projection basis variables which are the good variables to use here (see subsection \ref{purespindiel}). Actually, one can notice that the corresponding linear combinations (see (\ref{rel})) already appear in the previous equations, indicating the possible simplifications. One way to get them is to apply $\sigma$ on the equations and then project on the parallel and orthogonal parts\footnote{To do so, one has to know that $\sigma$ ``commutes'' with $\textrm{Re}()$ and $\textrm{Im}()$ (obvious), and more importantly, it commutes with $d()$, the exterior derivative, since the algebra of the manifold we consider has to be invariant under the projection (this is the same condition as the compatibility of the sources on this manifold, we will explain this in greater details with our solutions).}. This is another projection after the projection on real and imaginary parts and it gives much simpler equations. Note we have in each case $\sigma(H)=-H$. Using furthermore the projection conditions (\ref{projb1}), and (\ref{rel}), the SUSY conditions are simplified to:
\bea
\textrm{IIA}&:& d(\textrm{Im}(z))k_{\bot}=-g_s*F_4 \nn\\
& & d(\textrm{Im}(\Omega_2))k_{\bot}=k_{||}H \nn\\
& & -k_{||}d(\textrm{Im}(z))\w \textrm{Im}(\Omega_2)+\frac{1}{k_{||}}d(\textrm{Re}(\Omega_2)_{||})\w \textrm{Re}(z)-\frac{1}{k_{\bot}}H\w \textrm{Im}(z)=g_s*F_2 \nn\\
& & d(-\frac{1}{2}j^2\w \textrm{Im}(z))k_{\bot}-H\w(-k_{||}\textrm{Im}(z)\w \textrm{Im}(\Omega_2)+\frac{1}{k_{||}} \textrm{Re}(z)\w \textrm{Re}(\Omega_2)_{||})=-g_s*F_0 \nn\\
\nn\\
& & d(\textrm{Re}(z))=0 \nn\\
& & d(\textrm{Re}(\Omega_2)_{\bot})=k_{||}k_{\bot} \textrm{Re}(z)\w d(\textrm{Im}(z)) \nn\\
& & H\w \textrm{Re}(z)=-\frac{k_{\bot}}{k_{||}}d(\textrm{Im}(z)\w \textrm{Re}(\Omega_2)_{||}) \nn\\
& & d(j_{||}\w j_{\bot})=0 \nn\\
& & -\frac{1}{2}k_{||}d(j_{||}^2+j_{\bot}^2)+\frac{1}{k_{\bot}}\textrm{Re}(\Omega_2)_{\bot}\w \textrm{Re}(z)\w d(\textrm{Im}(z))=k_{\bot} H\w \textrm{Im}(\Omega_2) \nn\\
& & d(\textrm{Im}(\Omega_2)\w \textrm{Re}(\Omega_2)_{\bot})=0 \nn\\
& & H\w \textrm{Re}(z)\w \textrm{Im}(\Omega_2)=-H\w \textrm{Im}(z)\w \textrm{Re}(\Omega_2)_{||} \ ,
\eea
\bea
\textrm{IIB}&:& d(\textrm{Im}(\Omega_2))k_{\bot}=k_{||}H \nn\\
& & d(\textrm{Re}(\Omega_2)_{||})=k_{\bot}e^{i\theta}g_s*F_3 \nn\\
& & H\w \textrm{Re}(\Omega_2)_{||}=k_{\bot}e^{i\theta}g_s*F_1 \nn\\
\nn\\
\nn\\
& & d(\textrm{Re}(z))=0 \nn\\
& & d(\textrm{Im}(z))=0 \nn\\
& & \textrm{Re}(z)\w H=-\frac{k_{\bot}}{k_{||}}\textrm{Im}(z)\w d(\textrm{Re}(\Omega_2)_{\bot}) \nn\\
& & \textrm{Im}(z)\w H=\frac{k_{\bot}}{k_{||}}\textrm{Re}(z)\w d(\textrm{Re}(\Omega_2)_{\bot}) \nn\\
& & d(j_{||}\w j_{\bot})=0 \nn\\
& & -\frac{1}{2}k_{||}d(j_{||}^2+j_{\bot}^2)-\frac{1}{k_{\bot}} \textrm{Re}(z)\w \textrm{Im}(z) \w d(\textrm{Re}(\Omega_2)_{||})=k_{\bot} H\w \textrm{Im}(\Omega_2) \nn\\
& & -\textrm{Re}(z)\w H\w \textrm{Im}(\Omega_2)=\textrm{Im}(z)\w H\w \textrm{Re}(\Omega_2)_{\bot} \nn\\
& & \textrm{Im}(z)\w H\w \textrm{Im}(\Omega_2)=\textrm{Re}(z)\w H\w \textrm{Re}(\Omega_2)_{\bot} \ .
\eea

The final steps to get the SUSY conditions (\ref{SUSYA}) and (\ref{SUSYB}) are the following. One can first use the property derived in the next subsection, namely that in IIA/IIB there cannot be any $6$-form which is positive/negative under $\sigma$. This gives the automatic annihilation of the last equation of IIA and the two last equations of IIB, and the simplification of the definition of $F_0$ in IIA. Second, one can use the $SU(2)$ structure conditions, namely (\ref{com3}) to (\ref{com7}), to get some more simplifications.

\subsubsection{More use of the projection basis}\label{More}

In IIA/IIB we introduce on a six-dimensional manifold an $O6/O5$ plane. The $1$-form basis used is $(e^1, ..., e^6)$ and we choose the three/two internal dimensions of the $O6/O5$ along directions labeled $e^i_+$. The other three/four directions are labeled $e^i_-$. The $\pm$ are used in reference to the action of $\sigma$ on these forms: $\sigma(e^i_\pm)=\pm e^i_\pm$. We then deduce that any $i$-form $O_i$ can be decomposed naturally as $O_{i||}+O_{i\bot}$, which can only be written this way:
\bea
&& O_{1||}=\sum_{i} c^i_+\ e^i_+ \ ,\nn\\
&& O_{1\bot}=\sum_{i} c^i_-\ e^i_- \ ,\nn\\
&& O_{2||}=\sum_{i,j} c^{ij}_{||+}\ e^i_+\w e^j_+ + c^{ij}_{||-}\ e^i_-\w e^j_- \ ,\nn\\
&& O_{2\bot}=\sum_{i,j} c^{ij}_{\bot}\ e^i_+\w e^j_- \ ,\nn\\
&& O_{3\bot}=\sum_{i,j,k} c^{ijk}_{\bot+}\ e^i_+\w e^j_+\w e^k_- +  c^{ijk}_{\bot-}\ e^i_-\w e^j_- \w e^k_- \ ,\nn\\
&& ...
\eea

We can now show very easily that some conditions are automatically satisfied, or simplified, because we only have a limited number of $e^i_\pm$ in each theory. Especially, one can say that in IIA/IIB there cannot be any $6$-form which is positive/negative under $\sigma$, due to the number of $e^i_\pm$, and so we can get the automatic annihilation of some conditions. It is the case in the SUSY conditions given above.

\section{Discussion of some normalization with calibrated smeared sources}\label{cali}

In this appendix, we motivate the normalization condition (\ref{normVi}). From the work done on calibrations of supersymmetric sources \cite{Calib, KT}, we know that a calibrated source wrapping an internal $k$-dimensional cycle $\Sigma$ (in a $d$-dimensional internal space $M$), taken in a configuration without any flux pulled-back on it or world-volume flux $\mathcal{F}$, should satisfy the following condition:
\beq
\textrm{Im}(\Phi_2)|_{\Sigma}=\frac{|a|^2}{8} \sqrt{|det(P(G))|} \ d\sigma^1 \w \dots \w d\sigma^k \ ,
\eeq
where $\sigma^i$ are coordinates on $\Sigma$, $|det(P(G))|$ is the absolute value of the determinant of the pull-back on the source world-volume of the ten-dimensional metric $G$, and $\textrm{Im}(\Phi_2)$ is restricted to its components on $\Sigma$. With our ansatz (\ref{metricansatz}) for $G$, we get:
\beq
\textrm{Im}(\Phi_2)|_{\Sigma}=\frac{|a|^2}{8} e^{4A} V_{\Sigma}\ ,
\eeq
where $V_{\Sigma}$ is the volume form of $\Sigma$. Further, with our conventions and in the large volume limit, we get:
\beq
e^{3A-\phi} \textrm{Im}(\Phi_2)|_{\Sigma}=\frac{1}{8g_s} V_{\Sigma} \label{Vsig}\ ,
\eeq
where $e^{3A-\phi}$ should be understood as taken in the large volume limit.\\

The literature on calibrations introduces a current $j_{\Sigma}$, defined in our conventions as (the Mukai pairing was defined in (\ref{Mukai}))
\beq
\int_M \left\langle j_{\Sigma}, f \right\rangle=\int_{\Sigma} f \label{jsig}\ ,
\eeq
for a given form $f$ of $\Sigma$, and so one can introduce the one, $j_{\Sigma_i}$, associated to $e^{3A-\phi} \textrm{Im}(\Phi_2)|_{\Sigma_i}$ for a source $i$. This current is actually related to the source current appearing in the right-hand side of the BI. Indeed, we can write (up to some factors that we won't take into account)
\beq
(d-H\w)F=j_{Total}=\sum_{sources\ i} Q_i\ j_{\Sigma_i} \ ,
\eeq
$Q_i$ being considered as the RR charge. So $j_{\Sigma_i}$ corresponds to the density current, and can be written roughly as:
\beq
j_{\Sigma_i}\approx \delta^{d-k}(\Sigma)\ *V_{\Sigma_i} \ ,
\eeq
i.e. as a $\delta$ function to localize the source in its transverse directions, times the volume orthogonal to the cycle. Actually, the definition (\ref{jsig}) shows that a sign like the one given by $\lambda(f)$ is entering the game, because a Mukai pairing is used instead of a simple wedge product. Hence we choose\footnote{Note that we could multiply this expression for $j_{\Sigma_i}$ by $\frac{\int_{\Sigma_i} V_{\Sigma_i}}{\int_{M} V}$, a natural factor when considering (\ref{jsig}), which would make $j_{\Sigma_i}$ metric independent. This is one more example of positive factors which could be taken into account.}
\beq
j_{\Sigma_i}= \delta^{d-k}(\Sigma)\ *\lambda(V_{\Sigma_i}) \ .
\eeq

The smearing of the source corresponds to the idea that the source is not localized anymore in the transverse directions, or in other words, one doesn't see the $\delta$ function anymore, and so in this case, we write:
\beq
(d-H\w)F=\sum_{sources\ i} Q_i V^i,\qquad j_{\Sigma_i}=V^i \ ,
\eeq
and we should now have
\beq
V^i=*\lambda(V_{\Sigma_i})\ .
\eeq
Actually, one can show in our conventions that
\beq
\left\langle *\lambda(V_{\Sigma_i}), V_{\Sigma_i} \right\rangle \ =V \ ,
\eeq
(where $V$ is the internal space volume form). Hence, using (\ref{Vsig}) and the last result, we get to the following normalization condition in the large volume limit and for smeared sources:
\beq
\left\langle V^i, e^{3A-\phi} \textrm{Im}(\Phi_2) \right\rangle = \frac{1}{8g_s}\ V \ .
\eeq

We conclude with two remarks. First, this normalization could be refined, to take into account some forgotten factors like those appearing in the BI. But all these factors are positive, so they are not changing the sign of the charges, which is what matters in the end. Second, there are several ways to show that
\beq
\int_{M_6} \left\langle V^i, e^{3A-\phi} \textrm{Im}(\Phi_2) \right\rangle \ > 0 \ ,
\eeq
either by (\ref{jsig}), or by the derivation of the no-go theorem done this way in \cite{Scan}, hence the sign given by $\lambda(..)$ is indeed needed.

\section{Solutions with several O-planes}\label{reducset}

In this appendix, we are going to explain the arguments that allow to reduce the list of possible sets of manifolds/O-planes for an intermediate $SU(2)$ solution with several (non completely overlapping) orientifolds, as explained in subsection \ref{other}. Let us first consider the case of a type IIB solution (with an $O5$-plane). Using the same notations as in appendix \ref{More}, we introduce the natural notation for the $e^i$: $e^i_\pm$, defined by $\sigma(e^i_\pm)=\pm e^i_\pm$. For an $O5$-plane in a six-dimensional manifold, there are four $e^i_-$ and two $e^i_+$. To have an $O5$ source, it must first be compatible with the algebra of the manifold. The list given in \cite{Scan} indicate what are the possible compatible O-planes, so it must be part of it. Furthermore, one should have a non-trivial $F_3$ to see this source appear, and we recall that the BI for that flux ($d(F_3)$) gives the co-volume of this source\footnote{Note we consider here the simple case where there is no $-H\w F_1$ term, so the reasoning might not be the most general one. In IIA for $F_2$, we will not have this restriction since $-H\w F_0$ is an exact term.}. This co-volume is nothing but the wedge product of the four $e^i_-$. So there must be in the algebra of the manifold an $e^k$ such that $d(e^k)=e^i_-\w e^j_-$. This is actually a non-trivial requirement: we have to look for manifolds which have two O-planes being listed in the compatible O-planes, and satisfying this co-volume requirement. For instance, the nilmanifold $n\ 4.4$ has the following algebra: $(0,0,0,0,12,14+23)$. The $O5$ compatible are along $56$, $13$, and $24$. $56$ is the only one satisfying the co-volume requirement, so there can be at most one O-plane on this manifold. Note that solutions with an O-plane along $56$ are actually found in \cite{Scan} and \cite{KT}. Doing this systematic check, some manifolds are excluded.\\

We can add other criteria on the O-planes. For a type IIB solution, we have $\sigma(z)=-z$. So the O-planes must be orthogonal to it (the ``$z$ criteria''). The $z$ contains at least two distinct directions (otherwise $z\w \overline{z}$ would be zero which is forbidden) so the O-planes must be orthogonal to both. Furthermore, we have the SUSY condition $d(z)=0$, hence the O-planes must be at least orthogonal to two of the $0$ directions of the algebra. This ``$z$ criteria'' allows to discard all the $n\ 2.p$ nilmanifolds for instance, and also all the $s\ 1.p$ solvmanifolds.\\

There is another important criteria. Let us use the following notation: $z$ is at least along $e^1_-$ and $e^2_-$, there are two other $-$ directions noted $e^3_-$ and $e^4_-$, and the two $+$ directions are noted $e^1_+$ and $e^2_+$, with respect to the first O-plane. Each O-plane has to be orthogonal to $z$, so we can use the same notation for the second one: $z$ is at least along $e^1_-$ and $e^2_-$ which are $-$ directions for the second O-plane too. To have an intermediate $SU(2)$ structure solution, we must have a non-zero $\textrm{Re}(\Omega_2)_{||}$ (see the conditions). As explained in appendix \ref{More}, it is clear that $\textrm{Re}(\Omega_2)_{||}$ only has components on two-forms $e^i_+\w e^j_+$ or $e^i_-\w e^j_-$ but not on mixed $+$ and $-$. Because of the volume form condition given in (\ref{comb1}), and because $z$ is along $-$ directions, $\textrm{Re}(\Omega_2)_{||}$ is in (\ref{comb1}) the only form which can bring the $e^i_+$, so the pair $(e^1_+,e^2_+)$ has to be present in the decomposition of $\textrm{Re}(\Omega_2)_{||}$. For the solution to be compatible with both O-planes, $\textrm{Re}(\Omega_2)_{||}$ must be ``parallel'' under both projections, so it means that the pairs $(e^1_+,e^2_+)$ and $(e^i_-,e^j_-)$ of one O-plane must corresponds to such pairs for the other O-plane, and not to mixed $+$ and $-$ pairs. More precisely, as we do not want the O-planes to be completely overlapping, so not along the same two directions, we deduce that the pair $(e^1_+,e^2_+)$ for the first O-plane must be a $(e^k_-,e^l_-)$ pair for the second and vice-versa. But the two O-planes already share two $-$ directions given by $z$: $e^1_-$ and $e^2_-$. So $(e^1_+,e^2_+)$ of the first O-plane corresponds to the pair $(e^3_-,e^4_-)$ of the second and vice-versa. This just means the following ``direction criteria'': the two O-planes have to be along completely different directions, they cannot share one direction. These different directions are orthogonal to the $z$ directions, so not much possibility is left: in particular, $z$ is then ``only'' along two directions, and at most two non completely overlapping $O5$ are possible at the same time in an intermediate $SU(2)$ solution. This is exactly the case for our solutions: for instance for the first solution, the O-planes are along $45$ and $26$, and $z$ is along $1$ and $3$.\\

Applying carefully all these criteria, we find that the only remaining candidates are (with the a priori allowed directions of the O-planes in brackets):
$$n\ 3.3\ (45,16),\ n\ 3.6\ (25,46),\ n\ 3.9\ (25,46),\ n\ 3.13\ (45,26)$$
$$n\ 3.14\ (45,26),\ n\ 4.1\ (26,35,45),\ n\ 4.2\ (26,35,45),\ n\ 4.5\ (35,45,26,16),\ n\ 4.6\ (35,26)$$
$$s\ 2.2\ (14,23),\ s\ 2.4\ (14,15,24,25),\ s\ 2.5\ (13,14,23,24),\ s\ 2.6\ (14,23)$$
\beq s\ 3.1\ (14,15,16,24,25,26),\ s\ 3.3\ (13,14,23,24),\ s\ 4.1\ (14,15,16,24,25,26)\ .\eeq
Of course we find our solutions among them. Note that only some couples of the O-planes indicated are possible. If one wants to find solutions to the list of conditions (\ref{projb1}), (\ref{com3}), (\ref{com4}), (\ref{com5}), (\ref{com6}), (\ref{com7}), (\ref{comb1}) and (\ref{SUSYB}), one can use some symmetry properties to avoid testing all the possibilities. For instance, $n\ 4.2$ could a priori have a solution with the couples of O-planes $(26,35)$ and $(26,45)$. But its algebra, $(0,0,0,0,12,15)$ is clearly symmetric under the exchange of $3$ and $4$, so one can restrict the search to one of the two couples. The same goes for $n\ 4.1$ for instance by doing the change of variables ($e^3 \rightarrow e^4,\ e^4 \rightarrow e^3,\ e^2 \rightarrow -e^2,\ e^5 \rightarrow -e^5,\ e^6 \rightarrow -e^6$). In this way, the list of manifolds/couples of O-planes to test is limited to:
$$n\ 3.3\ (45,16),\ n\ 3.6\ (25,46),\ n\ 3.9\ (25,46),\ n\ 3.13\ (45,26)$$
$$n\ 3.14\ (45,26),\ n\ 4.1\ (26,35),\ n\ 4.2\ (26,35),\ n\ 4.5\ (35,26),\ n\ 4.6\ (35,26)$$
$$s\ 2.2\ (14,23),\ s\ 2.4\ (14,25),\ s\ 2.5\ (13,24),\ s\ 2.6\ (14,23)$$
\beq s\ 3.1\ (14,25) (15,24),\ s\ 3.3\ (13,24),\ s\ 4.1\ (14,25)\ .\eeq

Let us now consider the type IIA case, with $O6$ as sources. This gives in six dimensions three $e^i_-$ and three $e^j_+$. One can actually use the same kind of criteria. The ``co-volume'' criteria works the same with a non-trivial $F_2$: there must be in the algebra of the manifold an $e^k$ such that $d(e^k)=e^i_-\w e^j_-$. The ``$z$ criteria'' also works: $\textrm{Re}(z)$ is parallel to the O-plane, and its derivative is $0$, so the O-planes have to share at least one direction which gives a zero in the algebra. $\textrm{Im}(z)$ has to be orthogonal to the O-planes so they have to share at least one $e^i_-$. We recall that $\textrm{Re}(z)$ and $\textrm{Im}(z)$ are both non-zero and give at least two directions otherwise the volume form would be zero. So for each O-plane remain two $+$ and two $-$ directions. Can they share them? Considering exactly the same argument as before with $\textrm{Re}(\Omega_2)_{||}$, we get the following ``direction criteria'': the non completely overlapping O-planes share exactly one direction, the one given by $\textrm{Re}(z)$, and no other. This leads once again to the fact that at most two non completely overlapping $O6$ are possible at the same time in an intermediate $SU(2)$ solution. Applying all these criteria we get to the following reduced list:
$$n\ 3.9\ (235,346),\ n\ 3.10\ (136,235),\ n\ 3.11\ (136,235),\ n\ 3.15\ (235,346),\ n\ 3.16\ (136,235)$$
$$n\ 4.2\ (236,246,345),\ n\ 4.3\ (146,345),\ n\ 4.4\ (146,236,345),\ n\ 4.6\ (246,345),\ n\ 4.7\ (135,146,236,245)$$
$$s\ 2.2\ (135,245),\ s\ 2.5\ (136,146,236,246),\ s\ 2.6\ (146,236),\ s\ 3.2\ (146,145,256,356)$$
\beq s\ 3.3\ (136,146,236,246),\ s\ 3.4\ (145,246,346),\ s\ 4.1\ (145,156,256,245,146,246)\ .\eeq
Considering the symmetries, we get the following list to be tried:
$$n\ 3.9\ (235,346),\ n\ 3.10\ (136,235),\ n\ 3.11\ (136,235),\ n\ 3.15\ (235,346),\ n\ 3.16\ (136,235)$$
$$n\ 4.2\ (236,345),\ n\ 4.3\ (146,345),\ n\ 4.4\ (146,345),\ n\ 4.6\ (246,345),\ n\ 4.7\ (135,146) (135,236)$$
$$s\ 2.2\ (135,245),\ s\ 2.5\ (136,246),\ s\ 2.6\ (146,236),\ s\ 3.2\ (146,256)$$
\beq s\ 3.3\ (136,246),\ s\ 3.4\ (145,246),\ s\ 4.1\ (145,246)\ . \eeq
Of course we recover our solutions in these lists (they pass all the criteria).

\end{document}